\documentclass[default,iicol]{sn-jnl-tweak}

\usepackage{graphicx}
\usepackage{multirow}
\usepackage{amsthm}
\usepackage{mathrsfs}
\usepackage[title]{appendix}
\usepackage{xcolor}
\usepackage{textcomp}
\usepackage{manyfoot}
\usepackage{booktabs}
\usepackage{algorithm}
\usepackage{algorithmicx}
\usepackage{algpseudocode}
\usepackage{listings}
\usepackage{epsfig}
\usepackage{amsmath, amsfonts, amssymb, amsthm, amscd,booktabs, mathtools}
\usepackage{mathabx}
\usepackage{dsfont}
\usepackage{natbib}
\graphicspath{figures}
\usepackage[shortlabels]{enumitem}
\usepackage{hyperref}
\hypersetup{colorlinks,allcolors=black}
\usepackage{bookmark}
\usepackage{enumitem}
\usepackage{soul}
\usepackage[normalem]{ulem}

\hypersetup{
    colorlinks,
    linkcolor={blue!80!black},
    citecolor={green!50!black},
    urlcolor={magenta!80!black}
}

\hypersetup{
  pdfencoding=auto,
  unicode=true,
}

 \usepackage{manyfoot}

\usepackage{cleveref}
\usepackage{crossreftools}
\newtheorem{theorem}{Theorem}

\newtheorem{proposition}[theorem]{Proposition}

\theoremstyle{definition}

\newcommand{\cG}{\mathcal{G}}

\newcommand{\cI}{\mathcal{I}}

\newcommand{\cN}{\mathcal{N}}

\newcommand{\NN}{\mathbb{N}}

\newcommand{\RR}{\mathbb{R}}

\newcommand{\argmin}{\mathop{\mathrm{arg\,min}}}

\newcommand{\Var}{{\mathop{\mathrm{Var}}}}
\newcommand{\Bias}{{\mathop{\mathrm{Bias}}}}
\newcommand{\Cov}{{\mathop{\mathrm{Cov}}}}

\renewcommand{\bar}{\overline}

\renewcommand{\tilde}{\widetilde}

\newcommand\ddfrac[2]{\frac{\displaystyle #1}{\displaystyle #2}}

\newcommand{\pto}{\xrightarrow{\textup{p}}}
\newcommand{\dto}{\xrightarrow{\textup{d}}}

\newcommand{\far}{\mathtt{FAR}}

\renewcommand{\hat}{\widehat}

\usepackage{amsmath, amsfonts, amscd, amssymb}

\def\0{\mathbf{0}}
\def\1{\mathbf{1}}
\def\2{\mathbf{2}}

\newcommand{\mbbE}{\mathbb{E}}

\newcommand{\mbbP}{\mathbb{P}}

 \newcommand{\ind}{\perp\!\!\!\!\perp} 
 \newcommand{\norm}[1]{\left\lVert#1\right\rVert}
 \newcommand{\fnmr}{\mathtt{FRR}}
 \newcommand{\fmr}{\mathtt{FAR}}
 \newcommand{\hfnmr}{\widehat{\mathtt{FRR}}}
 \newcommand{\hfmr}{\widehat{\mathtt{FAR}}}
 
 \newcommand{\bhfnmr}{\widehat{\mathtt{FRR}}_b^*}
\newcommand{\bhfmr}{\widehat{\mathtt{FAR}}_b^*}
\newcommand{\bVar}{{\mathop{\mathrm{Var}}}^*}
\newcommand{\bCov}{{\mathop{\mathrm{Cov}}}^*}

\definecolor{editone}{HTML}{000000}

\newcommand{\editedinline}[1]{{\color{editone} #1}}

\definecolor{editone2}{HTML}{0000FF}
\definecolor{editone}{HTML}{000000}

\newcommand{\editedinlinetwo}[1]{{\color{editone2} #1}}

\newcommand{\titletext}
{\editedinline{Confidence Intervals for Error Rates in 1:1 Matching Tasks:\\ Critical Statistical Analysis and Recommendations}}

\begin{document}

\title[Confidence Intervals for Error Rates in 1:1 Matching Tasks]{\titletext}

\author*[1]{\fnm{Riccardo} \sur{Fogliato}}\email{fogliato@amazon.com}

\author*[2]{\fnm{Pratik} \sur{Patil}} \email{pratikpatil@berkeley.edu}

\author[3]{\fnm{Pietro} \sur{Perona}} \email{peronapp@amazon.com}

\affil*[1]{\orgname{AWS ML \& Engines}, \orgaddress{\country{USA}}}

\affil*[2]{\orgname{University of California, Berkeley}, \orgaddress{\country{USA}}}

\affil[3]{\orgname{AWS AI Labs}, \orgaddress{\country{USA}}}

\abstract{
Matching algorithms predict relationships between items in a collection. For example, in 1:1 face verification,  a matching algorithm predicts whether two face images depict the same person.  Accurately assessing the uncertainty of the error rates of such algorithms can be challenging when test data are dependent and error rates are low,  two aspects that have been often overlooked in the literature.
\editedinline{In this work, we review methods for constructing confidence intervals for error rates in 1:1 matching tasks.}
We derive and examine the statistical properties of these methods, demonstrating how coverage and interval width vary with sample size, error rates, and degree of data dependence with experiments on synthetic and real-world datasets.
\editedinline{Based on our findings, we provide recommendations for best practices for constructing confidence intervals for error rates in 1:1 matching tasks.}
}


\maketitle

\section{Introduction}
\label{sec:intro}

Accurately measuring system accuracy is essential for responsible design and deployment of automated systems~\citep{kearns2019ethical}. Accurate measurements aid in identifying suitable use cases for a system, guiding engineers towards enhancements, and helping stakeholders comprehend the system's strengths and limitations.  Nevertheless, the value of accuracy measurements is limited without considering their statistical uncertainty. While methodology for computing confidence intervals for classification tasks on independent data in well-established~\citep{brown2001interval}, it remains problematic for {\em matching tasks}.

\editedinline{To construct confidence intervals for the accuracy of algorithms used in 1:1 matching tasks,
using standard Wald intervals based on the Gaussian approximation 
of the maximum likelihood estimator may appear to be a viable approach~\citep{casella2021statistical, wasserman2004all}.}
However, this approach is problematic for the following two reasons:
\begin{enumerate}[\rm(M\arabic*)]
    \item
    \label{mot:low-error}
    \underline{\emph{Low error rates.}} When the 1:1 matching algorithm is highly accurate, as is the case, for instance, in  face recognition (FR) systems~\citep{grother2019face}, error rates are close to zero, which makes the Gaussian approximation inaccurate. Consequently, confidence intervals based  on this approximation may significantly under-cover the true error rates.
    \item
    \label{mot:sample-dependence}
    \underline{\emph{Sample dependence.}}
    When test sets are relatively small the pair-wise samples used in matching tasks may include the same item multiple times, e.g. the same face photograph may be used in multiple comparisons. This means that the samples are correlated. Therefore, using Wald intervals with variance estimated under the independence assumption is not suitable for this scenario.
\end{enumerate}
\editedinline{Our study focuses precisely on these issues.} It is worth mentioning that we are not the first to consider low error rates and sample dependence. Bootstrap procedures have been proposed in the FR literature to address sample dependence and have been widely used in empirical studies~\citep{bolle2004error, poh2007performance, wu2016impact, phillips2011distinguishing}. However, the development of these methods was based on heuristic arguments, and there has been limited discussion regarding their statistical guarantees,
such as their frequentist coverage.
For instance, it is well known that confidence intervals based on bootstrap
resampling can fail to achieve nominal coverage in various settings, 
meaning that the probability of the true parameter being contained in the intervals is lower than the desired rate.
This issue is prominent when accuracy/error metrics are close to the parameter boundary
(e.g., false acceptance rate is close to 0)
or when sample sizes are small.
These issues are particularly pertinent in intersectional analyses
within bias assessments~\citep{balakrishnan2020towards},
where the number of images available for certain demographic subgroups
is often limited.

\begin{figure}
    \centering
        \includegraphics[width=0.48\textwidth]{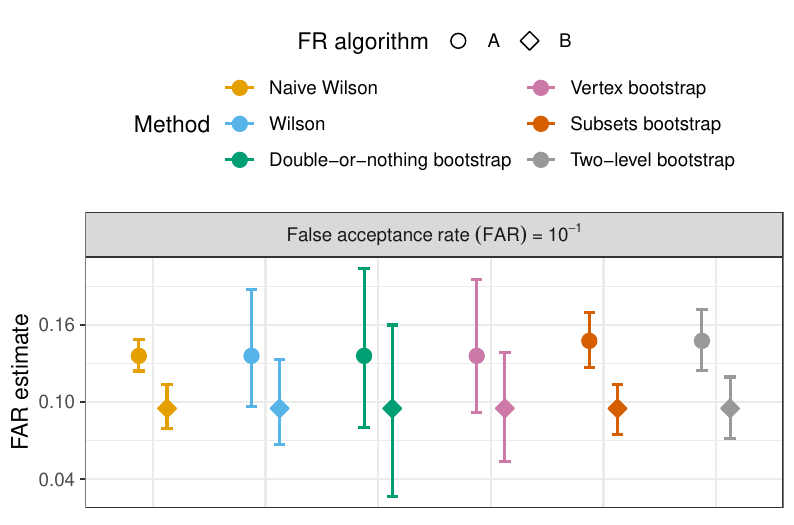}
    \caption{\editedinline{\textbf{Different methods for constructing confidence intervals 
    can lead to different conclusions due to miscoverage.} Six methods for computing estimates and corresponding 95\% confidence intervals on synthetic data for the false accept rate ($\fmr$) of two 1:1 matching algorithms (A and B) that have underlying equal accuracy ($\fmr=10^{-1}$). The data contains 50 groups, with 5 images each, and all pairwise comparisons are considered in the estimation of the error metric (details in \Cref{sec:experiments}). Dots and bars correspond to error estimates and corresponding confidence intervals. The naive Wilson, subsets bootstrap, and two-level bootstrap intervals may lead the practitioner to {\it erroneously} conclude that Algorithm A has inferior performance compared to Algorithm B -- while in our simulation they are equivalent. In our analysis and experiments we find that only Wilson intervals achieve nominal coverage in presence of low error rates \ref{mot:low-error} and sample dependence \ref{mot:sample-dependence}. Double-or-nothing and vertex bootstrap intervals also work well in settings characterized only by \ref{mot:sample-dependence}.}}
    \label{fig:width_intro}
\end{figure}

Different methods will yield confidence intervals with different widths. \Cref{fig:width_intro} shows such an example. Which is right? Without a thorough understanding of the statistical properties  of the various methods, it is difficult to determine which method is most appropriate  for a particular setting. It is unclear 
whether and when the constructed interval achieves  the desired nominal coverage. In light of these considerations,
our investigation is guided by the following fundamental question: 
\editedinline{\ul{\em Which methods should be used to construct confidence intervals 
for error metrics in 1:1 matching tasks?}}
We investigate methods with the primary aim of addressing the issues 
mentioned in \ref{mot:low-error} and \ref{mot:sample-dependence}.

\editedinline{Besides exploring analytically the properties of different methods, we carry out a thorough experimental investigation as well. We use both synthetic data and data coming from a real-life applications. Amongst many options we chose face verification, 
an important and sensitive application of Computer Vision~\citep{phillips2003face,phillips2018face,vangara2019characterizing,grother2019face}. 
Thus, we present a critical examination and analysis of methodologies for constructing confidence intervals for error rates in 1:1 matching tasks, and use face verification as a representative test application. Our findings are applicable across all 1:1 matching tasks, encompassing 1:1 speaker, fingerprint, and iris recognition, among others.}

\begin{figure*}[!ht]
    \centering
        \includegraphics[width=0.95\textwidth]{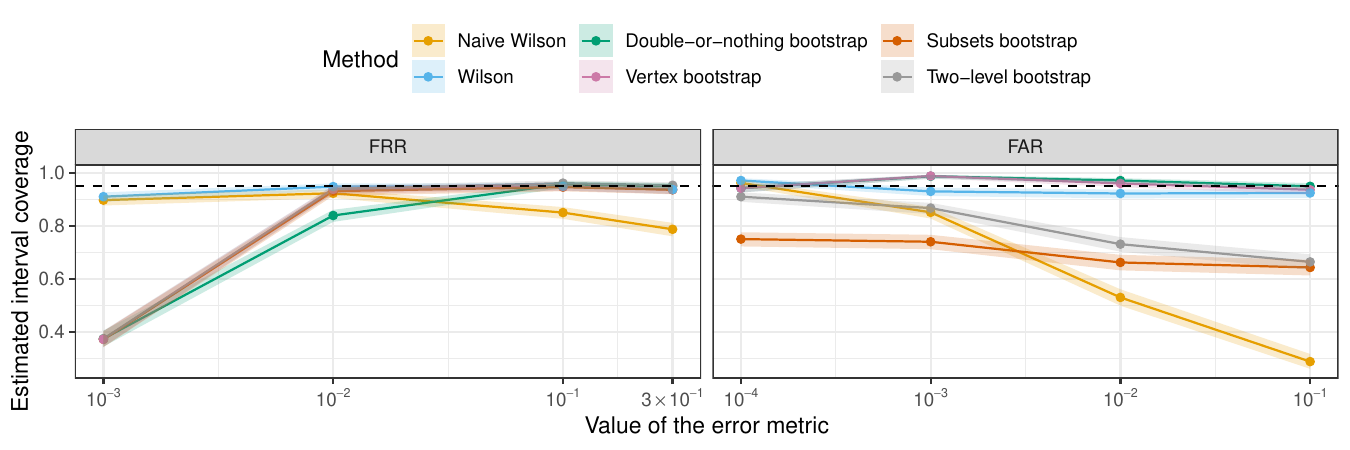}
    \caption{{\bf Estimated interval coverage of 95\% confidence intervals for $\fnmr$ and $\fmr$ on synthetic data in settings characterized by \ref{mot:low-error} and \ref{mot:sample-dependence}.} The data contains 50 identities and 5 instances (e.g., images) per identity. Lines and shaded regions indicate estimated coverage and corresponding 95\% confidence intervals respectively, for each method, across data replications. The dashed line indicates nominal coverage (95\%). The experimental setup is described in \Cref{sec:experiments}. \editedinline{Only Wilson's method (blue lines) guarantees accurate coverage across all experimental conditions.}} %
    \label{fig:fmr_coverage}
\end{figure*}

Our theoretical analysis and empirical investigation reveal that, 
although there is no ``one-size-fits-all'' solution, 
only certain methods consistently achieve coverage that is close to nominal. 
Some concrete examples are illustrated in Figure \ref{fig:fmr_coverage}. 
From the figure,
we observe that in case of the $\fnmr$ (false rejection rate), all nonparametric bootstrap methods significantly under-cover when $\fnmr$ is close to the parameter boundary, while parametric Wilson intervals that assume data independence under-cover for large values of $\fnmr$. 
In case of the $\fmr$ (false acceptance rate), the subsets and two-level bootstrap techniques fail to achieve nominal coverage at any level of the error metric, while the {\it naive Wilson} interval, where one one neglects to account for data dependence, shrinks with growing $\fmr$. The remaining three methods are more promising: Wilson interval that accounts for data dependence (which we will refer to simply as {\em Wilson} hereafter) always achieves nominal coverage, while the vertex and double-or-nothing bootstraps cover at the right level when the true error metrics are large. 
See \Cref{sec:methods} for a description of aforementioned methods. 

\paragraph{Summary of contributions} 
Our main contributions are as follows:\\

\begin{enumerate}[leftmargin=*]
\item 
\underline{\emph{Methods review.}}
We provide a review of two classes of methods for constructing confidence intervals for matching tasks, one based on parametric assumptions, and the other on nonparametric, resampling-based methods. The reviewed methods include the Wilson intervals without (naive version) and with variance adjusted for data dependence, subsets, two-level, vertex, and double-or-nothing bootstraps.
\item 
\underline{\emph{Theoretical analysis.}}
We present a theoretical analysis of the reviewed methods
with a focus on intervals for error rates that are computed at a fixed threshold. 
Our analysis includes statistical guarantees for coverage of the intervals 
and their width.

\item 
\underline{\emph{Empirical evaluation.}}
To compare the properties of confidence intervals for error rates at a fixed thresholds as well as of pointwise intervals for the ROC generated by the reviewed methods,
we conduct experiments on both synthetic and real-world datasets, namely on MORPH \citep{ricanek2006morph}.

\item
\editedinline{\underline{\emph{Software library.}} In addition, an open-source code library in both R and Python that implements the investigated methods is available at
\href{https://github.com/awslabs/cis-matching-tasks}{https://github.com/awslabs/cis-matching-tasks}}.

\item
\editedinline{\underline{\emph{A recommendation.}}
Based on these findings, we recommend \ref{rec:use} using Wilson intervals with variance adjusted for data dependence to address \ref{mot:low-error} and \ref{mot:sample-dependence}, or utilizing vertex and double-or-nothing bootstrap intervals in settings that exhibit only \ref{mot:sample-dependence}.}
\end{enumerate}

\paragraph{Paper outline}
In \Cref{sec:related-work},
we provide an overview of related work on confidence interval for clustered data. 
In \Cref{sec:setup},
we describe the problem setup. 
We focus on the balanced setting, where each individual present in the data has an equal number of images. 
In \Cref{sec:methods}, 
we describe the statistical properties of the methods in the balanced setting. 
In \Cref{sec:extensions}, we present extensions of these methods, including estimation in the unbalanced setting (where the number of instances can vary across individuals), 
strategies for 
constructing pointwise confidence intervals for the receiver operating characteristics (ROC) curve.  
In \Cref{sec:experiments},
we provide numerical evaluation of
different methods. 
In \Cref{sec:discussion},
we discuss merits and pitfalls of the different methods, as well as directions for future work. 
The Appendix to the paper contains proofs of the theoretical results
stated in the main paper, additional numerical experiments,
and other miscellaneous details.

\section{Related work}
\label{sec:related-work}

Our primary objective is to construct confidence intervals that achieve nominal coverage robustly for parameters close to the boundary of the parameter space \ref{mot:low-error} by handling dependent samples \ref{mot:sample-dependence}. The issue raised in \ref{mot:low-error} has been extensively studied by statisticians, while the issue mentioned in \ref{mot:sample-dependence} arises in the analysis of data such as time series, networks, surveys, dyadic, and panel data, among others. Consequently, confidence interval construction methods that handle sample dependence have been developed in Economics, Statistics, and the social sciences. In this section, we briefly review existing parametric and nonparametric methods proposed in these fields, as well as those introduced by the FR community.

\paragraph{Parametric methods} 
The most commonly used parametric confidence intervals under data independence
are Wald intervals, which rely on the asymptotic normality of the maximum
likelihood estimator.
However, this assumption may not hold in finite samples and thus this type of
interval may fail to achieve nominal coverage \citep{brown2001interval}.
One prominent example is the one of the binomial proportion (e.g., error rates of classification tasks under data independence) being equal to the
sample size, for which Wald-type intervals as well as bootstrap-based
intervals are degenerate. For this reason, in this setting the Wilson~\citep{wilson1927probable}, Agresti-Coull~\citep{agresti1998approximate},
and Jeffreys intervals are preferred. When the independence
assumption is violated, one can account for the dependency structure in the data
in the estimation of the Gaussian variance \citep{miao2004effect}. Variance
estimation methods have been explored in the context of dyadic data
\citep{ snijders1999non,fafchamps2007formation,cameron2011robust}. These approaches are discussed in Section
\ref{sec:methods}.

\paragraph{Nonparametric resampling methods in FR}%
Nonparametric resampling methods offer an alternative approach to constructing confidence intervals that does not rely on the asymptotic normality assumption of the target statistic. In the context of 1:1 face verification, \citet{bolle2004error} propose the subsets bootstrap, which consists of resampling at the level of the identities: If an identity is sampled, every comparison in the data that involves that identity is included in the bootstrap sample. 
However, as we demonstrate in Section \ref{sec:methods}, the dependence structure between the bootstrapped and original datasets may differ significantly, resulting in under-coverage of $\fmr$ intervals.  
In an attempt to address this issue, \citet{poh2007performance} propose a two-level bootstrap where resampling occurs at both the identity and individual score levels. In case of $\fnmr$ intervals, these two techniques are standard choices of block bootstraps, which have been discussed in the statistical literature \citep{davison1997bootstrap, field2007bootstrapping} and are widely used in practice. However, we found that none of the articles in the FR literature we reviewed discuss the statistical properties of these procedures, except for the recent work by \citet{conti2022assessing}. They propose resampling individual images and deriving confidence intervals from the resulting bootstrap distribution, which needs to be recentered around the metric value on the original dataset. They argue that, asymptotically, this distribution converges to the law of the target statistic.

\paragraph{Nonparametric resampling methods in Statistics and Economics} 
There are a number of nonparametric bootstrap and subsampling techniques 
available for conducting inference on dyadic data
\citep{bickel2011method, snijders1999non, green2022bootstrapping,
    bhattacharyya2015subsampling, menzel2017bootstrap, davezies2021empirical,
    cameron2015practitioner, mccullagh2000resampling}. See \citet{graham2020network} for a comprehensive
review. 
The majority of these approaches are intended to offer asymptotic guarantees, where the distribution of the conditional data mean follows a Gaussian distribution under specific circumstances. As a result, these methods mainly focus on the degree to which the bootstrap distribution approximates the first two moments of the underlying distribution of the target statistic.
One method that has been widely employed in the social sciences is the vertex
bootstrap proposed by \citet{snijders1999non}.
The procedure proposed by \citet{conti2022assessing} is similar in spirit, the main difference being that only the former swaps comparisons between the same image with a random sample taken from the set of comparisons present in the original data.
 In our work, we investigate both the vertex bootstrap and a related method, the double-or-nothing bootstrap, which has been studied in the context of exchangeable arrays, such as dyadic data \citep{owen2012bootstrapping,davezies2021empirical}. Both of these methods have the desirable property that, asymptotically, the first two moments of the bootstrap distribution match those of the distribution of the error metric estimators that we consider in this work. This is not the case for the subsets and two-level bootstraps. 

\paragraph{Parametric bootstrap methods} 
An alternative to nonparametric resampling methods comes in the form of parametric bootstrapping. For example, \citet{mitra2007statistical} fit a generalized linear mixed model and obtain credible intervals by sampling from the model's posterior predictive distribution. However, while mixed models are capable of handling network data dependency and have been widely studied 
\citep{hoff2002latent,hoff2021additive},
fitting these models on large datasets, such as those found in face recognition applications, can be challenging. For this reason, we exclude this type of inference from our analysis. An alternative Bayesian approach is to model the distributions of scores. This is done, e.g., by \citet{chouldechova2022unsupervised}, although their focus is on semi- and unsupervised estimation.

\section{Problem Setup}
\label{sec:setup}

In this section,
we describe the problem setup
and introduce our notation.
We consider a set of $G$ different identities
(we use the word ``identity'' commonly used in FR, 
where it means ``a specific person'')
denoted by $\cG$.
Each identity $i\in\cG$ has $M_i$ instances (e.g., face images) 
that are represented by embeddings $X_{(i, 1)}, \dots, X_{(i, M_i)} \in \RR^d$. 
For example, these embeddings are generated 
by a FR model and may be normalized.
We assume that
the embeddings follow a common probability law $Q$ on $\RR^{d}$
for all identities $i \in \cG$ and all $1\leq k\leq M_i$.
Furthermore,
if we consider a pair of instances $k$ and $l$ 
belonging to identities $i$ and $j$ from $\cG$, 
we assume that the embeddings $X_{(i, k)}$ and $X_{(j, l)}$ 
are independent when $i\neq j$. 

\editedinlinetwo{We will focus on binary classification tasks, 
where the goal is to classify a pair of instances
as belonging to the same identity (i.e., ``genuine'')
or different identities (i.e., ``impostor'').
This classification is done based on the distance (e.g., Euclidean distance)
between the embeddings and a threshold $t \in \RR$.
Specifically,
the pair of instances is classified as genuine
when
when $d(X_{(i, k)}, X_{(j, l)}) < t$,
and as impostor
when $d(X_{(i, k)}, X_{(j, l)}) \geq t$ for some distance function $d$.
For an identity $i$, when $k\neq l$, let 
$Y_{(i, k), (i, l)} = \mathds{1}\{d(X_{(i, k)}, X_{(i, l)}) \geq t\} \sim \text{Bernoulli}(\fnmr)$, 
where $\fnmr$ is termed False Non-Match Rate (FNMR) or False Rejection Rate ($\fnmr$). 
For different identities $i \neq j$, let $Y_{(i, k), (j, l)} = \mathds{1}\{d(X_{(i, k)}, X_{(j, l)}) <  t\} \sim \text{Bernoulli}(\fmr)$,
where $\fmr$ is termed False Match Rate (FMR) or False Acceptance Rate ($\fmr$).}
\editedinlinetwo{We are interested in estimating the parameters $\fnmr$ and $\fmr$ 
from the sample.}\footnote{\editedinlinetwo{The framework can be adapted to include conditioning on predefined attributes of identities, such as when it's already known which demographic groups certain identities belong to.}}\footnote{\editedinlinetwo{The framework can also apply to other losses such as cross-entropy. The principles and methods reviewed, including the bootstrap techniques, can be adapted or directly employed.}}

Apart from estimating these parameters,
we wish to construct
confidence intervals for them. 
There are two properties of the intervals 
one generally cares about:
one is coverage and the other is length.
Our primary focus is to construct
intervals with valid nominal coverage.
Formally, 
given a nominal coverage level of $1 - \alpha$
for some $\alpha \in (0, 1)$,
our goal is to construct confidence (rather than credible) intervals,
denoted by $\cI_{\fnmr}$ and $\cI_{\fmr}$, respectively,
for the two metrics $\fnmr$ and $\fmr$
that satisfy the following frequentist coverage guarantees: 
$\mbbP_Q(\fnmr\in \cI_\fnmr)\geq 1-\alpha$ and $\mbbP_Q(\fmr\in \cI_\fmr)\geq 1-\alpha$. 
Among intervals with the correct coverage,
shorter intervals are preferable.
In practice, however,
it may be difficult to achieve the exact coverage guarantee,
and thus, 
one might have to settle for an approximate guarantee.

We next consider estimators for $\fnmr$ and $\fmr$
in the form of empirical averages for the balanced setting, where $M_i = M$ for all $i \in \cG$. 
These point estimators lead to the confidence intervals
that we describe in Section \ref{sec:methods}. 
The estimation in the unbalanced setting, where the number of instances $M_i$ can vary across identities, is described in \Cref{sec:unbalanced_setting}.

\paragraph{Balanced setting}
Consider a sample 
where each of the $G$ identities available has $M$ instances. 
One can define natural empirical estimators of $\fnmr$ and $\fmr$ for identities $i,j\in\cG$ as follows: 
\begin{align}
    \bar{Y}_{ij}
    =
    \begin{dcases}
    \sum_{k=1}^M\sum_{\substack{l = 1,\\l \neq k}}^{M} \frac{Y_{(i, k), (i, l)}}{M(M-1)} & \text{ when } i=j,\\ 
    \sum_{k,l=1}^M \frac{Y_{(i,k), (j,l)}}{M^2} & \text{ when } i\neq j.
    \end{dcases}
\end{align}
Here, the estimator $\bar{Y}_{ij}$
measures the error metrics at the level of each identity.
Estimators of $\fnmr$ and $\fmr$ 
for the entire sample are then given by  
\begin{equation}\label{eq:estimators_balanced}
    \hfnmr = \sum_{i=1}^G \frac{\bar{Y}_{ii}}{G},
    \;
    \hfmr = \sum_{i=1}^G\sum_{\substack{j=1,\\ j\neq i}}^{G} \frac{\bar{Y}_{ij}}{G(G-1)},
\end{equation}
respectively. 
The type of confidence intervals for $\fnmr$ and $\fmr$
that we consider are based on these estimators.
It is easy to see that $\hfnmr$ and $\hfmr$ are unbiased estimators of $\fnmr$ and
$\fmr$ respectively, that is $\Bias(\hfnmr) = \mbbE[\hfnmr] - \fnmr = 0$ and $\Bias(\hfmr) = \mbbE[\hfmr] - \fmr = 0$. 
In addition, we have:
\begin{align}
    \Var(\hfnmr) &= \frac{1}{G}\Var(\bar{Y}_{11}), \label{eq:var_fnmr}\\
        \Var(\hfmr) &= 
        \frac{2}{G (G - 1)}
        \Var(\bar{Y}_{12}) \nonumber \\
        & \qquad +
        \frac{4 (G - 2)}{G (G - 1)}
        \Cov(\bar{Y}_{12}, \bar{Y}_{13})
        \label{eq:var_fmr}.
\end{align}
The variances will be of key interest throughout our discussion of the validity of the confidence intervals. 
While $\Var(\hfnmr)$ corresponds to the variance of $\hfnmr$ across identities, we observe that $\Var(\hfmr)$ will coincide with variance under data independence only when $\Cov(\bar{Y}_{12}, \bar{Y}_{13})= 0$. Thus, in general, the covariance terms will need to be accounted for 
in the construction of the confidence intervals. 

Our asymptotic analysis in Section \ref{sec:methods} will focus on the setting where $G$ grows while $M$ remains fixed. 
This is motivated by the observation that in FR applications the number of unseen identities is generally larger than the number of face images per identity. This kind of asymptotic analysis is also typical in prior studies on inference using clustered data \citep{field2007bootstrapping, cameron2015practitioner}.

\section{Methods description}
\label{sec:methods}

In this section, we describe parametric (\Cref{sec:methods_parametric}) and nonparametric resampling-based methods (\Cref{sec:methods_nonparametric}) for constructing confidence intervals for error rates in matching tasks with binary model predictions. Our focus will be on the balanced setting.
Methods extensions, including confidence intervals in the unbalanced setting, pointwise intervals for the receiver operating characteristic (ROC) curve, as well as protocol design strategies,  
can be found in \Cref{sec:extensions}. 
We will defer all the proofs of the theoretical statements 
to \Cref{sec:proofs}.

\subsection{Parametric methods}\label{sec:methods_parametric}

Parametric methods for constructing confidence intervals rely on assumptions made about the distribution of the target statistic. In \Cref{sec:related-work}, we have mentioned that Wald intervals are typically used for constructing intervals for statistics that are asymptotically normal under data independence, while other methods such as Wilson, Agresti-Coull, and Jeffreys have been explored specifically for binomial proportions. To derive intervals that have good coverage in the presence of data dependence, we need to characterize the asymptotic behavior of $\sqrt{G}(\hfnmr - \fnmr)$ and $\sqrt{G}(\hfmr - \fmr)$. The following proposition establishes a set of conditions under which these statistics are asymptotically normal.

\begin{proposition}
[Normality of scaled error rates]
\label{prop:normality}
Assume that $\lim_{G\rightarrow\infty}\Var(\sqrt{G}\hfnmr)=c_{\fnmr}$ and $\lim_{G\rightarrow\infty}\Var(\sqrt{G}\hfmr)=c_{\fmr}$ for some positive constants $c_\fnmr, c_\fmr$. Then,  
as $G \to \infty$, 
$\sqrt{G}(\hfnmr - \fnmr) \dto \cN(0,\Var(\bar{Y}_{11}))$ 
and 
$\sqrt{G}(\hfmr - \fmr) \dto \cN(0,4\Cov(\bar{Y}_{12}, \bar{Y}_{13}))$.
\end{proposition}

Since identity-level observations are assumed to be independent, 
the convergence in distribution of $\sqrt{G}\fnmr$ follows from an application of the central limit theorem. 
The case of the $\fmr$ follows from Proposition 3.2 in \citet{tabord2019inference}. 
The result in the proposition motivates 
the construction of confidence intervals 
based on the limiting distribution.

\paragraph{Construction of confidence intervals} 
The use of confidence intervals for binomial proportions in presence of dependent data was first proposed by \citet{miao2004effect}. 
For instance, Wald intervals in this setting 
take the form $\cI_{\fnmr}=[\hfnmr \pm z_{1 - \alpha/2} \sqrt{\Var(\bar{Y}_{11}) / G}]$ and $\cI_{\fmr}=[\hfmr \pm z_{1 - \alpha/2} \sqrt{\Cov(\bar{Y}_{12}, \bar{Y}_{13}) / G}]$, 
where $z_{1 - \alpha/2}$ corresponds to the $(1-\alpha/2)$-th quantile of the standard normal. 
From \Cref{prop:normality},
it then follows that
the intervals have the correct asymptotic coverage.
In practice, Wilson intervals are preferred as they achieve good coverage even in presence of small sample sizes \citep{brown2001interval}. 
The $1-\alpha$ {\em Wilson} confidence interval for $\fmr$, which assumes data dependence, is given by
\begin{align}
    \label{eq:modified-wilson-fmr}
    \cI_{\fmr}
    &=
    \Bigg[
    \frac{\hfmr\hat{N}_\fmr^* + \frac{1}{2}z_{1 - \alpha/2}^{2}}{\hat{N}_\fmr^* + z^{2}_{1 - \alpha/2}}  \pm \frac{z_{1 - \alpha/2} \sqrt{\hat{N}_\fmr^*}}{\hat{N}_\fmr^* + z_{1 - \alpha/2}^{2}}\nonumber\\& \quad  \cdot \sqrt{\hfmr(1-\hfmr) + z_{1 - \alpha/2}^{2}/(4\hat{N}_\fmr^*})
    \Bigg],
\end{align}
where $\hat{N}^*_\fmr =\max\{\hfmr(1-\hfmr)/\Var(\hfmr), \lfloor G/2 \rfloor\}$. 
The {\it naive Wilson} confidence interval, which assumes data independence, for $\fmr$ uses $\hat{N}_{\fmr}^* = G(G-1)M^2/2$.
The Wilson interval for $\fnmr$ is obtained by replacing $\hfmr$ with $\hfnmr$ and $N^*_{\fmr}$ with $N^*_\fnmr = \max\{\hfnmr (1 - \hfnmr)/ \Var(\hfnmr), G\}$, while its naive version employs $\hat{N}_{\fnmr}^* = G M(M-1)/2$. If Proposition \ref{prop:normality} holds, then the Wilson intervals will have the nominal coverage. 
As a side remark,
it is worth mentioning that the 95\% Wilson interval \eqref{eq:modified-wilson-fmr}
bears resemblance to a Wald interval that is calculated 
on a dataset with two successes and two failures appended.
For more details, see \citet{agresti1998approximate}.

It should be noted that the construction of these intervals 
relies on having knowledge of $\Var(\hfnmr)$ and $\Var(\hfmr)$. 
However, 
 thanks to Slutsky's theorem,
by replacing these variances
with their consistent estimators,
it is possible to use a modified version of Proposition \ref{prop:normality}
and certify the coverage of the resulting intervals.
In the following discussion, 
we will focus on constructing consistent estimators
of $\Var(\bar{Y}_{11})$ and $\Cov(\bar{Y}_{12}, \bar{Y}_{13})$.

\paragraph[Estimation of variances of FRR and FAR]{Estimation of $\Var(\hfnmr)$ and $\Var(\hfmr)$}
The variances in Proposition \ref{prop:normality} can be estimated using the following plug-in estimators:
\begin{align}
    \widehat{\Var}(\sqrt{G}\hfnmr) &= \frac{1}{G}\sum_{i=1}^G (\bar{Y}_{11} - \hfnmr)^2 \label{eq:var_fnmr_est}, \\
    \widehat{\Cov}(\bar{Y}_{12}, \bar{Y}_{13})
    &= \frac{1}{G (G-1) (G-2)} 
    \nonumber\\
    & \hspace{-12mm} \cdot
    \sum_{i = 1}^{G} 
    \sum_{\substack{j=1,\\j \neq i}}^{G} 
    \sum_{\substack{k=1\\k \neq j, i}}^{G}
    (\bar{Y}_{ij} - \hfmr)(\bar{Y}_{ik} - \hfmr).
    \label{eq:varest_plugin_covfmr} %
\end{align}
The estimator in \eqref{eq:var_fnmr_est} is the standard variance estimator under data independence. 
The estimator in \eqref{eq:varest_plugin_covfmr} is employed for $\Cov(\bar{Y}_{12}, \bar{Y}_{13})$. 
However, in finite samples, when $\Var(\bar{Y}_{12})\gg \Cov(\bar{Y}_{12}, \bar{Y}_{13})$, the individual variance terms in \eqref{eq:var_fmr} may dominate. In that case, we may want to employ the following estimator of $\Var(\bar{Y}_{12})$:
\begin{gather}
    \widehat{\Var}(\bar{Y}_{12}) = \frac{1}{G(G-1)}\sum_{i=1}^G\sum_{\substack{j=1\\j\neq i}}^{G} (\bar{Y}_{ij}-\hfmr)^2,
\end{gather}
and then plug in the estimators above %
into the variance expression \eqref{eq:var_fmr}.
That is, 
we use:
\begin{align}\label{eq:varest_plugin}
     &\widehat{\Var}(\sqrt{G}\hfmr) = \nonumber \\
     &
      \frac{2}{G-1}
      \widehat{\Var}(\bar{Y}_{12})
      +
      \frac{4 (G-2)}{G-1} \widehat{\Cov}(\bar{Y}_{12}, \bar{Y}_{13}).
\end{align}
This estimator is a special case of the robust variance estimator proposed by \citet{fafchamps2007formation} in the context of dyadic regression.
The following proposition states the convergence in probability of these estimators to the target parameters. 
\begin{proposition}
[Consistency of plug-in variance estimators]
\label{prop:consistency_estimators_variance} 
Consider the variance estimators 
$\widehat{\Var}(\sqrt{G}\hfnmr)$
and
$\widehat{\Var}(\sqrt{G}\hfmr)$
defined in \eqref{eq:var_fnmr_est} and \eqref{eq:varest_plugin}, respectively.
Then, 
as $G \to \infty$,
    $\widehat{\Var}(\sqrt{G}\hfnmr) \pto \Var(\bar{Y}_{11})$,
    and $\widehat{\Var}(\sqrt{G}\hfmr)
    \pto 4\Cov(\bar{Y}_{12}, \bar{Y}_{13})$.
\end{proposition}

An alternative way to estimate $\Var(\sqrt{G}\hfmr)$ is 
by using the following jackknife estimator:
\begin{multline}
\label{eq:varest_jackk}
     \widehat{\Var}_{JK}(\sqrt{G}\hfmr)
     =  \frac{(G-2)^2}{G} \\ \cdot \sum_{i=1}^G (\hfmr_{-i} - \hfmr)^2  - 2\frac{\widehat{\Var}(\bar{Y}_{12})}{G-1}.
\end{multline}
Here, we have defined $\hfmr_{-i} = (G-1)^{-1}(G-2)^{-1}\sum_{j=1}^G \sum_{k=1,k\neq j}^{G} \bar{Y}_{jk}\mathds{1}(\{j\neq i\}\cap  \{k\neq i\})$. 
It turns out that the plug-in and jackknife estimators produce exactly the estimates.
This is formalized in the following proposition.
\begin{proposition}
[Equivalence of plug-in and jackknife variance estimators]
\label{prop:equivalence_jackk_pi}
    Consider the estimators $\widehat{\Var}(\sqrt{G}\hfmr)$ and $\widehat{\Var}_{JK}(\sqrt{G}\hfmr)$  defined in equations \eqref{eq:varest_plugin} and \eqref{eq:varest_jackk}, respectively.
    It holds that
    $\widehat{\Var}(\sqrt{G}\hfmr) = \widehat{\Var}_{JK}(\sqrt{G}\hfmr)$. 
\end{proposition}
The equivalence between the two estimators follows from the work of \citet{graham2020network}. The implementations of the two methods have similar computational costs, as they both involve $O(G^3)$ operations. 
Moreover, the estimators can be rewritten by using a multiway clustering decomposition, as outlined in Proposition 2 of \citet{aronow2015cluster}. 
The implementation of this method is available in existing software packages 
in both R \citep{zeileis2020various} and Python \citep{seabold2010statsmodels}.

\subsection{Resampling-based methods}\label{sec:methods_nonparametric}

We now consider an alternate and popular class of methods, 
confidence intervals constructed by bootstrap resampling.
Bootstrap confidence intervals employ the so-called bootstrap
distribution of the statistic of interest, 
which is obtained by resampling with replacement 
from the original data, the statistic of interest. 
In the percentile bootstrap method, 
the interval is based on the percentiles of this distribution \citep{diciccio1996bootstrap}. 
For instance, 
let $\{\bhfmr\}_{b=1}^B$ be the bootstrap distribution 
and let $\bhfmr$ be the $\fmr$ estimated on the $b$-th bootstrap sample (i.e., resampled dataset). 
A $1-\alpha$ confidence interval for $\fmr$ is given by $[\hfmr^*_{(\left\lfloor B(\alpha/2)\right\rfloor)}, \hfmr^*_{(\left\lceil B(1-\alpha/2)\right\rceil)}]$.
Below we discuss nonparametric resampling techniques that can be used to obtain the bootstrap distribution. 
The asymptotic coverage properties of the intervals constructed via the bootstrap depend on the mean and variance of the bootstrap distribution, hence we focus on these properties. In our subsequent discussion, we denote with $\mbbE^*$ and $\Var^*$ the expectation and variance conditional on the original sample. 
\Cref{tab:bootstrap_prop} summarizes the statistical properties of the bootstraps that will be reviewed in the current section. 

\begin{table}[!ht]
\centering
\caption{Overview of asymptotic bias of the variance of bootstrapped error rates for the methods reviewed. All bootstrapped estimators have unbiased first moments.}\label{tab:bootstrap_prop}
\resizebox{\columnwidth}{!}{%
\begin{tabular}{l|c|c}
\toprule
 \editedinline{\textbf{Bootstrap}} & $\Bias(\Var^*(\sqrt{G}\bhfnmr))$ & $\Bias(\Var^*(\sqrt{G}\bhfmr))$  
 \\
 \midrule
  Subsets & $\approx 0$ & $< 0$   \\
 Two-level & $\approx 0$ & $< 0$   \\
  Vertex & $\approx 0$ & $\approx 0$ \\
 \begin{tabular}{@{}c@{}}Double-or \\ \; -nothing\end{tabular} & $\approx 0$ & $\approx 0$ \\
 \bottomrule
\end{tabular}
}
\end{table}

\paragraph{Subsets bootstrap}

At a fundamental level, the naive bootstrap resamples individual comparisons at the level of either identities or instances (of identities). However, ignoring the dependence structure present in the data can lead to significant undercoverage, as seen in the naive Wilson method. The subsets bootstrap \citep{bolle2000evaluation} attempts to modify the conventional bootstrap by incorporating some of the dependency into the resampling process. Specifically, this bootstrap involves resampling with replacement at the identity level $G$ times in each iteration. If the $i$-th identity is drawn in the $b$-th repetition, then all comparisons involving that identity are included in the bootstrap sample. More precisely, let the multinomial vector $(W_1, \dots, W_G) \sim \text{Multinomial}(\mathcal{G}, (G^{-1}, \dots, G^{-1}))$ such that $\sum_{i=1}^G W_i = G$. Then, we calculate:
\begin{gather}\label{eq:metrics_subsets}
\hfnmr_b^* = \sum_{i=1}^G \frac{W_i \bar{Y}_{ii}}{G}, \; 
\bhfmr = \frac{\sum_{\substack{i, j=1\\j \neq i}}^G W_i \bar{Y}_{ij}}{G(G-1)}.
\end{gather}
By including all observations corresponding to a resampled identity, this bootstrap should better approximate the true distribution of $\hfnmr$ and $\hfmr$ than the conventional bootstrap. Unfortunately, in a balanced setting, this procedure will underestimate the variance of $\hfmr$, as the following proposition demonstrates.

\begin{proposition}
[Bias of subsets bootstrap estimators]
    \label{prop:bootstrap_var_subsets}
    For the subsets bootstrap,
    we have
    $\Bias(\bhfnmr) = 0$, and $\Bias(\bhfmr) = 0$. 
    In addition, 
    we have $\Bias(\Var^*(\sqrt{G}\bhfnmr)) = - \Var(\hfnmr)$, and $\Bias(\Var^*(\sqrt{G}\bhfmr)) = - \Var(\hfmr) -  \{\Var(\bar{Y}_{12}) + 3(G-2)\Cov(\bar{Y}_{12}, \bar{Y}_{13})\}/(G-1)$. 
\end{proposition}

Deriving the unbiasedness of the $\fnmr^*_b$ variance is straightforward, while proving the same for $\fmr_b^*$ requires more intricate analysis. The proposition shows that either taking bootstrap samples of size $G-1$ or rescaling the bootstrap metrics by $\sqrt{G(G-1)^{-1}}$ provide estimates whose distribution has unbiased variance for $\fnmr$. However, for $\fmr$, there is a significant negative bias if $\Cov(\bar{Y}_{12}, \bar{Y}_{13}) > 0$.

\paragraph{Two-level bootstrap} 

The two-level bootstrap is an attempt to address 
the undercoverage issue of the subsets bootstrap 
by employing two stages of resampling \citep{poh2007performance}. 
In the first stage, 
we use the subsets bootstrap,
while in the second stage 
we employ a naive bootstrap to resample with replacement 
the instance comparisons belonging to the data subsets 
obtained in the first stage. 
In other words, 
after drawing 
$(W_1, \dots, W_G)\sim \text{Multinomial}(\mathcal{G}, (G^{-1}, \dots, G^{-1}))$, 
we compute 
\begin{align}
    \hspace{-2mm}
    \hfnmr_b^* =  \frac{\sum_{i=1}^G W_i \bar{Y}^*_{ii}}{G},
    \;
    \bhfmr = \frac{\sum_{\substack{i, j=1\\j\neq i}}^G W_i \bar{Y}_{ij}^*}{G(G-1)}.
\end{align}
Here,
$\bar{Y}^*_{ii}$ 
and 
$\sum_{\substack{j=1, j\neq i}}^G \bar{Y}_{ij}^*$ 
are obtained by applying a naive bootstrap on each resampled data subset. 
By applying the law of total variance,
we can show that 
$\Bias(\Var^*(\sqrt{G}\hfnmr_b^*)) = -\Var(\hfnmr) + (1 - 2/[M(M-1)])\Var(Y_{(1, 1), (1, 2)}) + O(M^{-3})$. 
This implies that, 
after rescaling the estimates, 
the bootstrap may produce excess variation in $\fnmr$ computations. 
The derivation of the $\fmr$ variance follows a similar strategy.

\paragraph{Vertex bootstrap} 

An alternative resampling procedure
is the vertex bootstrap, 
which is commonly used for inference on networks in the social sciences \citep{snijders1999non}. 
This method involves resampling with
replacement at the level of the identities $G$ times, and then considering all comparisons between the resampled identities.   
In case of the $\fnmr$, this method is equivalent to the subsets bootstrap. 
For $\fmr$ computations, 
the comparisons between instances belonging to the same identity are swapped with $\hfmr$; note that the original version of this bootstrap swaps it with a random sample from all comparisons.   
That is, for the $b$-th bootstrap sample, we take $(W_1, \dots, W_G)\sim \text{Multinomial}(\mathcal{G}, (G^{-1}, \dots, G^{-1}))$ and obtain $\hfnmr_b^*$ as in expression \eqref{eq:metrics_subsets}, while
for $\bhfmr$, we use:
\begin{align}
    \label{eq:hfrr-vertex}
    \bhfmr &= \sum_{i, j=1}^G W_i\bigg[ \frac{(W_i - 1) \hfmr \mathds{1}(i=j)}{G(G-1)} \nonumber\\& \qquad \qquad \qquad + \frac{W_j \bar{Y}_{ij} \mathds{1}(i\neq j) }{G(G-1)}\bigg].
\end{align}

\begin{proposition}
[Bias of vertex bootstrap estimators]
\label{prop:bootstrap_var_vertex}
    For the vertex bootstrap, 
    we have $\Bias(\bhfmr) = 0$, and $\Bias(\Var^*(\sqrt{G}\bhfmr)) = [4(G-2)/G^3 + O(G^{-3})]\Var(\bar{Y}_{12}) - [28/[G(G-1)] + O(G^{-3})]\Cov(\bar{Y}_{12}, \bar{Y}_{13})$. %
\end{proposition}
By comparing  \Cref{prop:bootstrap_var_subsets} and \Cref{prop:bootstrap_var_vertex},
one observes that in large samples, 
the expected $\fmr$ bootstrap variance of the vertex bootstrap 
is closer to the true variance than the subsets bootstrap. 
However, in finite samples, 
the vertex bootstrap overestimates the variances of the individual observations. 
As we will observe in the experiments in \Cref{sec:experiments}, 
this behavior can cause the bootstrap to achieve a coverage rate 
that is higher than nominal coverage.

\paragraph{Double-or-nothing bootstrap} 
The double-or-nothing bootstrap has been proposed 
in the context of separately exchangeable arrays 
\citep{owen2012bootstrapping}, 
and it is a natural approach for analyzing matching tasks. 
In each iteration of this bootstrap procedure, 
we sample weights 
$W_i\stackrel{\text{iid}}{\sim} \text{Uniform}\{0,2\}$ 
for each identity  $i\in\mathcal{G}$, 
and then we compute the estimates $\hfnmr_b^*$ and $\hfmr_b^*$
as follows:
\begin{gather}\label{eq:db_estimators}
\begin{aligned}
    \hfnmr_b^* &= \frac{\sum_{i=1}^G W_i \bar{Y}_{ii}}{\sum_{i=1}^G W_i},\\
    \bhfmr &= \frac{\sum_{\substack{i, j=1\\j\neq i}}^G W_iW_j\bar{Y}_{ij}}{\sum_{\substack{i, j=1\\j\neq i}}^G W_iW_j}.
\end{aligned}
\end{gather}

\begin{proposition}
[Bias of double-or-nothing bootstrap estimators]
\label{prop:bootstrap_var_doublenothing}
    For the double-or-nothing bootstrap,
    we have $\Bias(\bhfnmr) = 0$, and $\Bias(\bhfmr) = 0$. 
    In addition, 
    we have  $\Bias(\Var^*(\sqrt{G}\hfnmr_b^*))= - \Var(\hfnmr)$ and $\Bias(\Var^*(\sqrt{G}\bhfmr))=[G-1]^{-1}\{- (4G-2)/G \Var(\hfmr) + 4\Var(\bar{Y}_{12})\}$. 
\end{proposition}
Thus, 
the $\fnmr$ estimates obtained through subsets, vertex, and double-or-nothing bootstraps share similar properties. 
However, when computing $\fmr$, this procedure, like the vertex bootstrap,
tends to overestimate the variances of the individual identity-level comparisons.

\section{Practical considerations}
\label{sec:extensions}

In Section \ref{sec:methods}, 
we described the construction
of confidence intervals in the simplified balanced setting
and analyzed their properties.
In this section, 
we will provide an overview practical considerations
related to the implementation of these methods.
Specifically, we will discuss how the reviewed methods 
can be extended and applied in the unbalanced setting. 
We will then address the construction of 
pointwise confidence intervals for the ROC curve. 
In \Cref{sec:protocol_design},
we also cover the design of the protocols for the estimation of error rates 
and the associated uncertainty levels on large datasets.

\subsection{Handling unbalanced datasets}\label{sec:unbalanced_setting}

In many datasets, the number of instances (say, face images) 
varies across identities (say, different people). 
We now demonstrate how the proposed methods can be applied 
to construct confidence intervals in this setting. 
Let
$M_i\in\mathbb{N}$ denote a random variable 
representing to the finite number of instances 
belonging to the $i$-th identity. 
\editedinline{We assume that $M_i\stackrel{\text{i.i.d.}}{\sim} L$ 
for $i\in \mathcal{G}$ and some probability law $L$ on natural numbers $\NN$.} 
We consider the following natural estimators of the error metrics:
\begin{align}\label{eq:unbalanced_metrics}
\hfnmr &= \frac{\sum_{i=1}^G \widetilde{M}_{i} \bar{Y}_{ii}}{\sum_{i=1}^G \widetilde{M}_i}, 
\\
\hfmr &= \frac{\sum_{i=1}^G \sum_{j=1,j \neq i}^{G} M_iM_j \bar{Y}_{ij} }{\sum_{i=1}^G \sum_{j=1,j \neq i}^{G}M_i M_j},
\end{align}
where we have defined $\widetilde{M}_i = M_i (M_i-1)$. 
Unlike in the balanced setting, 
these estimators are only unbiased as $G \to \infty$. 
The expressions of their variances are slightly more involved compared to \eqref{eq:var_fnmr} and \eqref{eq:var_fmr}, and will be discussed below. 

\paragraph{Parametric methods}

The construction of the parametric-based confidence intervals
of the form \eqref{eq:modified-wilson-fmr}
in the balanced setting
can be easily extended  
to the unbalanced setting once estimators of $\Var(\sqrt{G}\hfnmr)$ and $\Var(\sqrt{G}\hfmr)$ are available. 
To obtain an estimator for $\Var(\sqrt{G}\hfnmr)$,
we can apply the Delta method that yields:
\begin{multline}
    \label{eq:var-fnmr-unbalanced}
     \Var (\sqrt{G}\hfnmr) =  \dfrac{\Var(\tilde{M}_1 \bar{Y}_{11})}{\mbbE[\tilde{M}_1]^2} \\- 2 \dfrac{\mbbE[\tilde{M}_1 \bar{Y}_{11}]\Cov(\tilde{M}_1 \bar{Y}_{11}, \tilde{M}_1)}{\mbbE[\tilde{M}_1]^3} \\ + \dfrac{\mbbE[\tilde{M}_1\bar{Y}_{11}]^2\Var(\tilde{M}_1)}{\mbbE[\tilde{M}_1]^4}. %
\end{multline}
Note that when the number of instances per identity is constant, 
the expression in \eqref{eq:var-fnmr-unbalanced}
reduces to $\Var(\bar{Y}_{11})$, which corresponds to the variance of $\hfnmr$ in \eqref{eq:var_fnmr} in the balanced setting. 
Unlike the balanced setting,
however,
using plug-in estimators for the various terms in \eqref{eq:var-fnmr-unbalanced} may result in negative estimates of $\Var(\sqrt{G}\hfnmr)$. 
One way around this is to rely on a different plug-in estimator. 
For this derivation, 
we assume that $M_i$ is independent of $Y_{(j, k), (l, p)}$ for any of the indices (even $i=j$ or $i=l$), i.e., the number of instances available for each identity is independent of whether the model classification is correct.\footnote{This is a simplifying assumption that may not always hold true. For instance, in datasets containing mugshots like MORPH, individuals who have been arrested more frequently 
could be more identifiable because their facial images are more up-to-date.} Then, appealing to the Delta method and this independence assumption, we obtain
\begin{equation}
    \label{eq:var-fnmr-unbalanced-alternate}
    \Var (\sqrt{G}\hfnmr)  %
     = \frac{\mbbE[\widetilde{M}_1^2(\bar{Y}_{11} - \far)^2]}{\mbbE[\widetilde{M}_1]^2}.
\end{equation}
The expression in \eqref{eq:var-fnmr-unbalanced-alternate} provides a simple way to estimate the $\fnmr$ variance when the independence assumption holds. 

To derive the plug-in estimators for $\Var(\sqrt{G}\hfmr)$,
we use similar arguments. 
Under the independence assumption described above, 
the Delta method yields
\begin{multline*}
    \Var (\sqrt{G}\hfmr) 
    = 
    \frac{2}{G - 1}
    \frac{\mbbE\left[M_1^2M_2^2(\bar{Y}_{12} - \fmr)^2\right]}{\mbbE[M_1]^4}\\
    + 
    \frac{4 (G - 2)}{(G - 1)}
    \frac{\mbbE\left[ M_1^2M_2M_3(\bar{Y}_{12}\bar{Y}_{13} - \mbbE[\bar{Y}_{12}\bar{Y}_{13}])\right]}{\mbbE[M_1]^4}.
\end{multline*}
Once we have the estimators $\widehat{\Var}(\hfnmr)$ and $\widehat{\Var}(\hfmr)$ for $\Var(\hfnmr)$ and $\Var(\hfmr)$ respectively,
we can construct Wilson confidence intervals
using the recipe described in Section \ref{sec:methods_parametric}.

\paragraph{Resampling-based methods}

Adapting the resampling-based methods for interval construction
in the unbalanced setting is rather straightforward,
similarly to the confidence intervals based on parametric methods.
Since we have assumed that the $M_i$'s are i.i.d., 
the methods will operate in the same way as in the balanced setting, 
with the only exception being that the metric computations follow \eqref{eq:unbalanced_metrics}. 
In other words, 
the resampling is performed at the identity level,
regardless of the number of instances $M_i$ for each identity. 
According to the following proposition, 
the subsets, vertex, and double-or-nothing bootstrap variances 
asymptotically converge to the target parameter in case of the $\fnmr$.

\begin{proposition}
[Consistency of bootstrap estimators for $\fnmr$]
\label{prop:unbalanced_var_boot}
    Under the unbalanced setting,
    as $G \to \infty$,
    $\Var^*(\sqrt{G}\hfnmr^*_b) - \Var(\sqrt{G}\hfnmr)\stackrel{p}{\rightarrow}0$,
    where $\hfnmr^*_b$ is the $\fnmr$ estimate of the b-th subsets, vertex, or double-or-nothing bootstrap sample. 
\end{proposition}

This result also indicates that 
the bootstrap methods may be a suitable alternative
for estimating $\Var(\hfnmr)$ 
instead of relying on the previously described plug-in estimator. 
Note that we have not talked about $\fmr$
in \Cref{prop:unbalanced_var_boot}.
Proving the consistency for $\fmr$ is a more intricate task as it involves computing the variance of non-independent terms. 
Thus, we do not pursue it in this paper.

Lastly, it is worth making a note on the scenario
where the number of instances available for each identity
is fixed instead of being random.
In this situation,
the variance computations undergo slight modifications.
For instance, when computing the variance for the $\fnmr$,
we have $\Var(\hfnmr) = \sum_{i=1}^G\widetilde{M}_i^2\Var(\bar{Y}_{i} \mid \widetilde{M}_i)/(\sum_{i=1}^G\widetilde{M}_i^2)$,
where $\widetilde{M}_i$ is a fixed quantity. 
Moreover, when applying the bootstrap method in this context,
it is essential to resample conditioning on $\widetilde{M}$.

\subsection{Pointwise intervals for ROC curves}\label{sec:pointwise_roc}

In this section, we focus on the construction of pointwise confidence intervals for ROC curves, i.e., intervals for error metrics such as $\fnmr$@$\fmr$. While there is a wide range of techniques available \citep{fawcett2004roc, macskassy2005pointwise}, we will limit our discussion to a few strategies 
that have proven effective in previous work and in our own experiments. 

\paragraph{Parametric methods} 
To construct $1-\alpha$ pointwise confidence intervals for the ROC,
we can use the Wilson method, as well as other parametric methods such as Wald intervals, as follows.
First, 
we first compute a $1-\alpha_\fmr$ interval for $\fmr$, 
and we denote the lower and upper bounds of this interval as
$\hfmr_{\mathrm{lb}}$ and $\hfmr_{\mathrm{ub}}$.
Intuitively, these intervals contain $\fmr$ with high probability. 
We then estimate $1-\alpha$ confidence intervals for $\fnmr$ at the thresholds $t$ that yield $\hfmr_{\mathrm{lb}}$ and $\hfmr_{\mathrm{ub}}$. 
The resulting $\fnmr$@$\fmr$ interval is given by the region between the minima and maxima of the union of the two intervals. If the Wilson method is used, all $\fnmr$ intervals computed on $\fmr$ values within $[\hfmr_{\mathrm{lb}}, \hfmr_{\mathrm{ub}}]$ will be nested within this region. Thus, as long as $\alpha_{\fmr}$ is small, %
we should expect the resulting intervals to be conservative. 
In practice, we have found that even large values of $\alpha_\fmr$ may yield intervals whose coverage is close to nominal. 
Therefore, the parameter $\alpha_\fmr$ should be calibrated 
to the specific sample to avoid severe over- or under-coverage. 

\paragraph{Nonparametric resampling-based methods}

An alternative approach is to employ the nonparametric methods,
such as bootstrapping techniques, 
which we have discussed in the previous sections. 
In this approach,
we first obtain several ROC curves via some bootstrapping methods, such as the double-or-nothing or the vertex bootstraps. 
We then used these curves to construct the confidence intervals 
for $\fnmr$@$\fmr$. 
For example, in the vertical averaging technique 
\citep{fawcett2004roc, macskassy2005pointwise}, 
one computes $\hfnmr^*_b$@$\fmr$ for each curve and then via the percentile bootstrap obtains the interval $[\hfnmr^*_{(\lfloor B\alpha/2\rfloor)}$@$\fmr, ~\hfnmr^*_{(\lceil B(1-\alpha/2)\rceil )}$@$\fmr]$. 
However, as alluded to before,
the main issue with the bootstrap methods
is the interval under-coverage for error metrics close to the parameter boundary. This issue can be mitigated by imposing smoothness assumptions.
For instance,
instead of using the empirical ROC, we can estimate the ROC curve parametrically (e.g., with the widely used binormal model) or nonparametrically with kernels instead of using its empirical estimator \citep{krzanowski2009roc}.

\section{Empirical evaluation}
\label{sec:experiments}
\label{sec:empyrical}

We will first present the experiments on synthetic data 
in the balanced setting, followed by experiments on
on MORPH in the unbalanced setting. 
Additional experiments and results, including those on pointwise confidence intervals for the ROC, 
are reported in \Cref{sec:experiments_supplement} of the Appendix. 

\begin{figure*}[ht]
    \centering
        \includegraphics[width=0.95\textwidth]{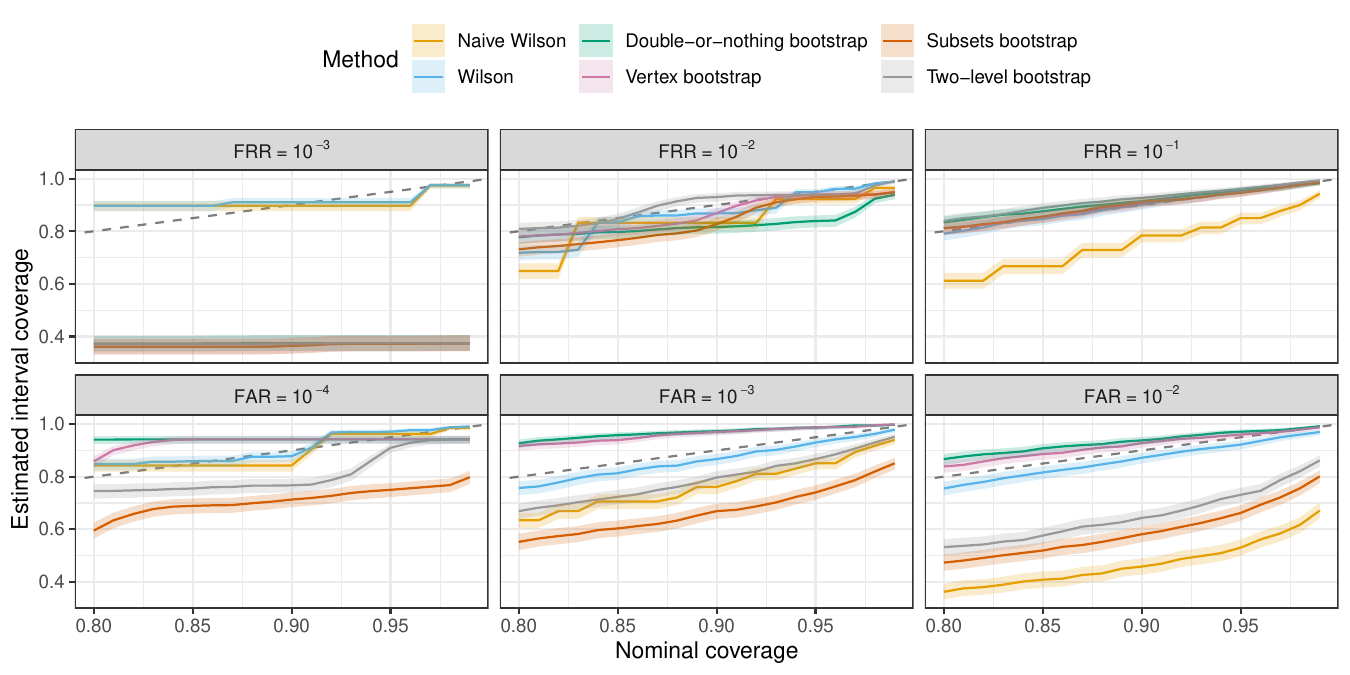}
    \caption{{\bf Estimated interval coverage versus nominal coverage for $\fnmr$ and $\fmr$ on synthetic data.} Data contain $G=50$ identities with $M=5$ instances each. Colored lines and shaded bands indicate estimated coverage computed on $10^3$ independent data replications and corresponding 95\% naive Wilson intervals for the coverage respectively. Ideally, estimated coverage would coincide with nominal coverage (black dashed line).}
    \label{fig:balanced_coverage}
\end{figure*}

\subsection{Experiments on synthetic data}
\label{sec:synthetic}

We consider the balanced setting with $G$ identities and $M=5$ instances for each identity. The embedding of the $k$-th image in the $i$-th identity are defined as: $X_{i, k}=\beta_i + \epsilon_{i, k}$, where $\beta_i, \epsilon_{i,k}\in\mathbb{R}^{128}$ with $\beta_i(d)\stackrel{\text{iid}}{\sim} \text{Exponential}(1)$ and $\epsilon_{i,k}(d)\stackrel{\text{iid}}{\sim} N(0, 5)$ for $1\leq d\leq 128$. 
We then define 
\begin{gather*}
    Y_{(i, k), (j, l)} = \begin{dcases} \mathds{1}\left(\norm{\tilde{X}_{i, k} - \tilde{X}_{j, l}}_2> t\right) & \text{ if } i=j,\\ \mathds{1}\left(\norm{\tilde{X}_{i,k}-\tilde{X}_{j,l}}_2\leq  t)\right) &  \text{ if } i\neq j,\end{dcases}
\end{gather*}
where we denote by $\tilde{X}_{i, k}=X_{i,k}/\norm{X_{i,k}}_2$
(and $\norm{\cdot}_2$ denotes the Euclidean norm). 
Here, $t > 0$, $i,j\in \cG$, and $1\leq k,l\leq M$, leaving $Y_{(i,k), (i,k)}$ undefined.
The error metrics estimation follows the description of Section \ref{sec:setup}.
The thresholds $t$ that yield the target error metrics, which are the underlying true parameters, were computed by resampling large datasets ($G=2\cdot 10^{2}$, $M=10$). Coverage and average width of the intervals were then estimated by repeating the described sampling process $10^2$ or $10^3$ times.

In \Cref{fig:balanced_coverage}, 
we compare estimated and nominal interval coverage 
for the methods discussed in Section \ref{sec:methods}
using synthetic data with $G=50$. 
We can derive three key takeaways, 
which we hinted when discussing \Cref{fig:fmr_coverage} in \Cref{sec:intro}.  

First, when $\fmr$ is far from $0$ (e.g., $\fmr=10^{-2}$ in this example), the Wilson intervals, vertex, and double-or-nothing bootstrap intervals achieve coverage close to nominal coverage. 
In contrast,
the naive Wilson, subsets, and two-level bootstrap intervals are too narrow and under-cover. 
Our empirical analysis confirms this finding, 
where we observed that only the naive Wilson intervals suffer 
from under-coverage when $\fnmr=10^{-2}$ or $10^{-1}$. 
By contrast, the two-level bootstrap tends to slightly over-cover.

Second, when $\fmr=10^{-3}$, the vertex and double-or-nothing bootstraps overestimate the variance of the $\fmr$ 
and thus produce intervals that are too large to be useful. Wilson intervals achieve coverage close to the nominal level, whereas the remaining intervals under-cover.

Third, when error metrics are small, actual coverage does not scale linearly with nominal coverage for any of the methods. The use of the bootstrap is most problematic in case of $\fnmr=10^{-3}$, as its distribution often results in a point mass at $0$ and thus leads to the observed severe under-coverage. 
Although the issue is somewhat mitigated in case of larger (relatively to the sample size) error metrics such as $\fmr=10^{-4}$, 
the bootstrap still may not achieve nominal coverage. 

\Cref{fig:fmr_coverage} and \Cref{fig:balanced_coverage} provide additional insights into the relationship between the two terms in \Cref{eq:var_fmr}, specifically $\Var(\bar{Y}_{12})$ and $\Cov(\bar{Y}_{12}, \bar{Y}_{13})$, respectively. Notably, as the $\fmr$ moves away from the parameter boundary, the ratio $\Cov(\bar{Y}_{12}, \bar{Y}_{13}) / \Var(\bar{Y}_{12})$ also increases.
This phenomenon is linked to a more pronounced under-coverage of the naive Wilson intervals and a less pronounced over-coverage of the vertex and double-or-nothing bootstrap intervals.

\begin{figure*}[ht]
    \centering
    \includegraphics[width=0.95\textwidth]{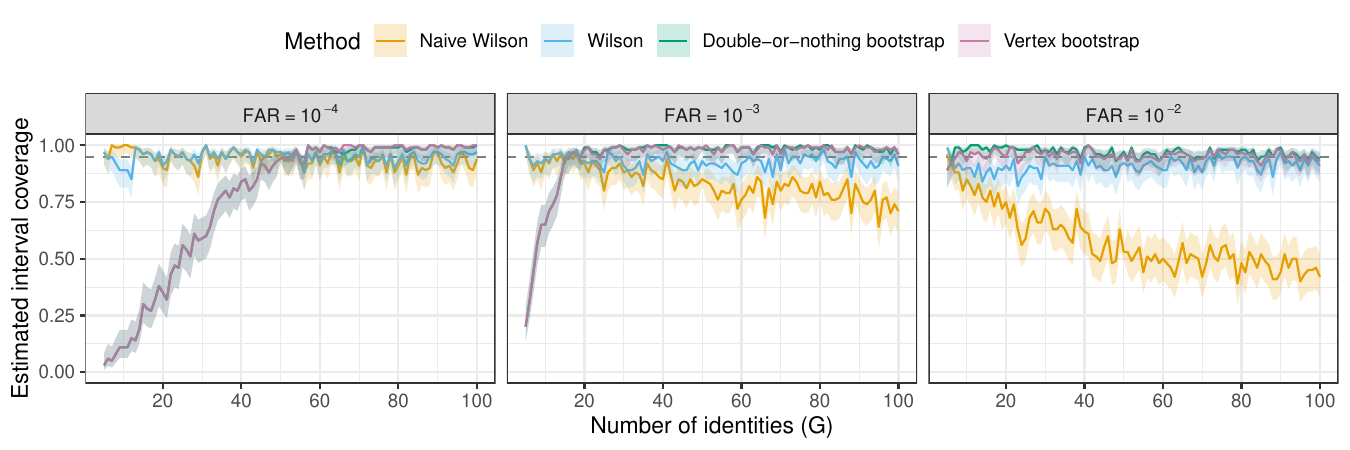}
    \caption{{\bf Estimated coverage of 95\% confidence intervals for $\fmr$ versus sample size on synthetic data}. Data contain $G$ identities (horizontal axis) with $M=5$ instances each.}\label{fig:fmr_coverage_vs_samplesize_balanced}
\end{figure*}

More generally, the findings from our study highlight the crucial relationship 
between the sample size and the magnitude of target error metric. \Cref{fig:fmr_coverage_vs_samplesize_balanced} provides additional insights
by depicting how the coverage of 95\% confidence intervals for $\fmr$ varies for $5\leq G\leq 100$. We observe that the coverage of naive Wilson intervals decreases with an increasing sample size, as the covariance terms become the leading factor in $\Var(\hfmr)$. Wilson intervals always cover approximately at the right level. In the case of the vertex and double-or-nothing bootstraps, they under-cover when $G$ is small and tend to over-cover for larger values of $G$, as can be observed for $G\geq 50$ (corresponding to 31k distinct comparisons) and $G\geq 20$ (5k comparisons) in case of $\fmr=10^{-4}$ and $\fmr=10^{-3}$, respectively. However, in line with our theoretical analysis, these intervals eventually achieve nominal coverage as $G$ keeps increasing, as demonstrated by the case of $\fmr=10^{-4}$.

\subsection{Experiments on MORPH}\label{sec:experiment_morph}

The MORPH dataset~\citep{ricanek2006morph} licensed for commercial use comprises approximately 400k mugshots images of 65k distinct individuals. As the number of images available for each individual varies, our estimation challenge is in the unbalanced setting. To expedite computations, we limited the number of face images 
per individual to 10.
Using \texttt{dlib}'s face recognition model through \texttt{DeepFace} \citep{king2009dlib, serengil2020lightface},
we extracted the 128-dimensional embeddings for the face images.
We then split the data in half and a large random sample of images from one half was employed to estimate the thresholds that yield the target $\fnmr$ and $\fmr$ using the Euclidean norm of the differences between the embeddings in the verification task.  
Construction of the confidence intervals was performed on the other half of the data. For this step, we generated datasets by randomly resampling without replacement $G$ identities and considering all pairwise comparisons between images corresponding to those identities. Estimation of error metrics and interval construction followed the method descriptions in Section \ref{sec:unbalanced_setting}. 

\begin{figure*}[!ht]
    \centering
    \includegraphics[width=0.95\textwidth]{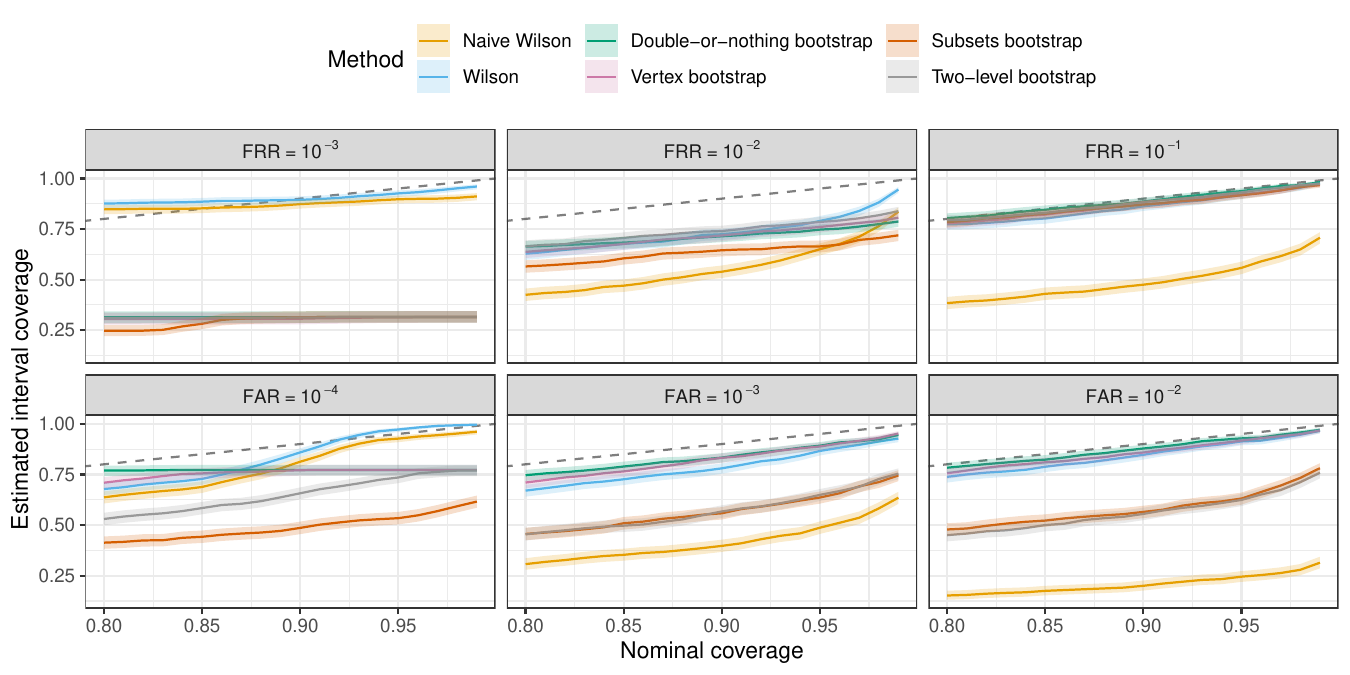}
    \caption{{\bf Estimated interval coverage versus nominal coverage for $\fnmr$ and $\fmr$ on MORPH.} Samples were generated by resampling $G=50$ identities from the original dataset without replacement.
    }\label{fig:morph_results}
\end{figure*}

Here, we will only focus on the methods that have produced the most promising results on synthetic data. Therefore, we exclude the subsets and two-level bootstraps but retain the naive Wilson as a baseline. \Cref{fig:morph_results} illustrates how the estimated coverage of confidence intervals for $\fmr$ and $\fnmr$ of these methods vary with nominal coverage on MORPH data with $G=50$, where different identities can have different numbers of images. The behavior of the intervals somewhat mirrors our observations on synthetic data.
More specifically, when the error metrics are close to zero ($\fmr=10^{-4}$ and $\fnmr=10^{-3}$), the double-or-nothing and vertex bootstrap intervals significantly under-cover, while the Wilson intervals perform better in this regard, although their actual coverage does not scale linearly with nominal coverage. For larger error metrics, the naive Wilson intervals are too narrow. In the case of $\fnmr$, all intervals under-cover when $\fnmr=10^{-2}$, while coverage is close to nominal when $\fnmr=10^{-1}$. For $\fmr$, all intervals tend to cover approximately at the nominal level when $\fmr=10^{-3}$ or $10^{-2}$.

\editedinlinetwo{Despite our theoretical guarantees on Wilson's method, we notice that in some cases there is undercoverage. This may happen for two reasons. First, our analysis is based on the assumption that observations are independent across identities or individuals. This assumption may not hold. For example, in scenarios where the same background, the same camera, or the same lighting are used for all mugshots taken only on a particular day, sequential dependence across identities may occur. Second, estimating the true values of the metrics of interest is challenging when using limited amounts of data.  As a result, the estimated interval coverage in the figures may not fully mirror the actual coverage in relation to the real value of the metric depicted in the figures.}
\section{Conclusions and Recommendations}
\label{sec:discussion}

 We aimed to provide guidelines for practitioners on how to compute confidence intervals for their experimental results. \editedinline{To this end, we explored the popular methods for constructing confidence intervals for error metrics in 1:1 matching tasks and evaluated their properties empirically and theoretically. } 
 Based on our findings:

\begin{enumerate}[\rm(R\arabic*)]
    \item 
     \label{rec:use} \ul{We recommend the use of Wilson intervals with adjusted variance}.
    They generally achieve coverage close to the nominal level. For large error metrics relative to sample size, the vertex and double-or-nothing bootstrap methods can be considered as good alternatives. 
    \item
     \label{rec:notuse}\ul{We strongly advise against using naive Wilson intervals, subsets, and two-level bootstrap techniques}.
     They fail to achieve nominal coverage and may lead to incorrect inferences.
\end{enumerate}

Our recommendations are especially relevant when test datasets are small-to-medium size, where all pairwise comparisons between instances are used in the computation of error metrics. On massive datasets non-overlapping sample pairs may be used, and data dependence may play a lesser role in the estimation of error metrics.

\editedinline{Our study is limited to  1:1 matching tasks. Computing confidence intervals for 1:N matching tasks is left open and will be the focus of future work.
Concepts and insights presented here will likely serve as a useful starting point towards that goal.}

\section*{Acknowledgments}

The authors would like to thank 
Mathew Monfort and Yifan Xing 
for the insightful discussions 
and valuable feedback on the paper. 
\editedinline{The anonymous reviewers and the associate editor are also gratefully acknowledged for their constructive feedback that helped improved the clarity of the paper.}

\bibliographystyle{apalike}
\bibliography{bibliography}

\begin{appendices}

\crefalias{section}{appendix}
\crefalias{subsection}{appendix}
\crefalias{subsubsection}{appendix}

\newcommand{\removelinebreaks}[1]{%
      \def\\{\relax}#1}
\def\titleRLB{\removelinebreaks{\titletext}}

\onecolumn

\begin{center}
\Large
{
Supplementary Material for \\
``\titleRLB''
}
\end{center}

\vspace{10mm}

\editedinline{
This document acts as a supplement to the paper 
``\titleRLB.''
}
The supplement is organized as follows. 

\medskip

\begin{enumerate}[label=(\Alph*)]
    \item 
    In \Cref{sec:proofs},
    we provide proofs of all the theoretical claims
    in the main paper.
    \begin{enumerate}
        \item[(1)]
        \Cref{sec:proofs_parametric}
        contains proofs
        for parametric methods
        in \Cref{sec:methods_parametric}.
        \item[(2)]
        \Cref{sec:proofs_nonparametric}
        contains proofs
        for resampling-based methods
        in \Cref{sec:methods_nonparametric}
        \item[(3)]
        \Cref{sec:proofs_unbalanced}
        contains proofs
        for unbalanced datasets
        in \Cref{sec:unbalanced_setting}.
    \end{enumerate}
    \smallskip
    
    \item In \Cref{sec:protocol_design}, we describe protocol design strategies (i.e., sampling) for the estimation of error rates and their associated uncertainty on large datasets. 
    \smallskip
    
    \item
    In \Cref{sec:experiments_supplement},
    we provide additional numerical experiments,
    supplementing those in \Cref{sec:experiments}.
\end{enumerate}

\section{Proofs of theoretical results}
\label{sec:proofs}

\subsection{Proofs for parametric methods in \Cref{sec:methods_parametric}}
\label{sec:proofs_parametric}

\subsubsection{Proof of \Cref{prop:normality} (normality of scaled error rates)}

As explained in the main paper,
because identity-level observations are assumed to be independent, 
the case of $\fnmr$ in Proposition \ref{prop:normality} 
follows from applying the central limit theorem. 
The case of the $\fmr$ follows from 
Proposition 3.2 in \citet{tabord2019inference}.

\subsubsection{Proof of Proposition \ref{prop:consistency_estimators_variance}
(consistency of plug-in variance estimators)}

The convergence in probability 
of $\widehat{\Var}(\sqrt{G}\hfnmr)$ to $\Var(\bar{Y}_{11})$ 
simply follows from an application of the weak law of large numbers.
In the following, we will show the convergence 
in probability
of $\widehat{\Var}(\sqrt{G} \hfmr) - (G - 1)^{-1}(2\Var(\bar{Y}_{12}) + 4(G-2)\Cov(\bar{Y}_{12}, \bar{Y}_{13}))$ to $0$. 
We begin by recalling the estimator:
\begin{align}
    \widehat{\Var}(\sqrt{G}\hfmr) 
    &= \frac{2}{G-1}\left[  \widehat{\Var}(\bar{Y}_{12}) + 2(G-2) \widehat{\Cov}(\bar{Y}_{12}, \bar{Y}_{13})\right],
\end{align}
where the components
$\widehat{\Var}(\bar{Y}_{12})$
and
$\widehat{\Cov}(\bar{Y}_{12}, \bar{Y}_{13})$
are defined as:
\begin{align}
    \widehat{\Var}(\bar{Y}_{12}) 
    &= \frac{1}{G(G-1)}\sum_{i=1}^G \sum_{j=1, j\neq i}^G (\bar{Y}_{ij} - \hfmr)^2,\\
    \widehat{\Cov}(\bar{Y}_{12}, \bar{Y}_{13})
    &= \frac{1}{G(G-1)(G-2)}\sum_{i=1}^G\sum_{\substack{j=1\\j\neq i}}^G\sum_{\substack{k=1\\k\neq i, j}}^G (\bar{Y}_{ij} - \hfmr) (\bar{Y}_{ik} - \hfmr).
\end{align}

We want to show that, 
as $G\rightarrow\infty$, 
$\widehat{\Var}(\bar{Y}_{12}) \pto \Var(\bar{Y}_{12})$,
and
$\widehat{\Cov}(\bar{Y}_{12}, \bar{Y}_{13}) \pto \Cov(\bar{Y}_{12}, \bar{Y}_{13})$. If these conditions are verified, then $ \widehat{\Var}(\sqrt{G}\hfmr)  -  \Var(\sqrt{G}\hfmr) \pto 0$ by Slutsky's theorem. 

\paragraph[Consistency of variance estimator]{Consistency of $\widehat{\Var}(\bar{Y}_{12})$} 
By Chebyshev's inequality,
we have
\begin{equation}\label{eq:chebychev_variance}
    \mathbb{P}(\vert \widehat{\Var}(\bar{Y}_{12}) -\Var(\bar{Y}_{12}) \vert \geq t)\leq \frac{\mbbE\left[ \left( \widehat{\Var}(\bar{Y}_{12}) -\Var(\bar{Y}_{12})\right)^2 \right]}{t^2},
\end{equation}
for any $t > 0$. 
We will now bound the numerator of \eqref{eq:chebychev_variance}. 
Decompose the numerator into:
\begin{gather}\label{eq:bias_var_decomposition}
    \mbbE\left[ \left(\widehat{\Var}(\bar{Y}_{12})) -  \Var(\bar{Y}_{12})\right)^2 \right] 
    = 
    \underbrace{\Var(\widehat{\Var}(\bar{Y}_{12}))
    \vphantom{\left( \mbbE\left[\widehat{\Var}(\bar{Y}_{12})\right] - \Var(\bar{Y}_{12}) \right)^2}
    }_{\textbf{Term 1}}
    + 
    \underbrace{\left( \mbbE\left[\widehat{\Var}(\bar{Y}_{12})\right] - \Var(\bar{Y}_{12}) \right)^2}_{\textbf{Term 2}}. 
\end{gather}
We will show below that both the two terms 
on the right-hand side of \eqref{eq:bias_var_decomposition} are $O(G^{-1})$.

\framebox{\textbf{Term 1}}
The first term in \eqref{eq:bias_var_decomposition} is equal to 
\begin{multline}\label{eq:bias_var_decomposition_first_term}
    \Var(\widehat{\Var}(\bar{Y}_{12})) = \frac{1}{G(G-1)}\bigg[ 2\Var((\bar{Y}_{12} - \hfmr)^2) + 4(G-2)\Cov((\bar{Y}_{12} - \hfmr)^2, (\bar{Y}_{13} - \hfmr)^2)\\
    + (G-2)(G-3)\Cov((\bar{Y}_{12} - \hfmr)^2, (\bar{Y}_{34} - \hfmr)^2) \bigg]. 
\end{multline}
It is easy to see that all terms are $O(G^{-1})$ or of smaller order. 

\framebox{\textbf{Term 2}}
The second term on the right hand side of \eqref{eq:bias_var_decomposition} is equal to
\begin{align}\label{eq:bias_var_decomposition_second_term}
    \left[ \mbbE\left[ \widehat{\Var}(\bar{Y}_{12}) \right] - \Var(\bar{Y}_{12})\right]^2 = - \left[ \frac{\Var(\bar{Y}_{12}) + 4(G-2)\Cov(\bar{Y}_{12}, \bar{Y}_{13})}{G(G-1)}\right]^2 = O(G^{-2}). 
\end{align}
The consistency of 
$\widehat{\Var}(\bar{Y}_{12})$ then follows by combining the results in \eqref{eq:bias_var_decomposition_first_term} and \eqref{eq:bias_var_decomposition_second_term} with the inequality in \eqref{eq:chebychev_variance}.

\paragraph[Consistency of covariance estimator]{Consistency of $\widehat{\Cov}(\bar{Y}_{12}, \bar{Y}_{13})$}
By Chebyshev's inequality,
\begin{gather}\label{eq:chebychev_cov}
    \mathbb{P}(\vert \widehat{\Cov}(\bar{Y}_{12}, \bar{Y}_{13}) -\Cov(\bar{Y}_{12}, \bar{Y}_{13}) \vert \geq t)\leq \frac{\mbbE\left[ \left( \widehat{\Cov}(\bar{Y}_{12}, \bar{Y}_{13}) -  \Cov(\bar{Y}_{12}, \bar{Y}_{13})\right)^2 \right]}{t^2},
\end{gather}
for any $t > 0$.
We now proceed to bound the numerator of \eqref{eq:chebychev_cov}.
Note that
\begin{align}\label{eq:cov_decomposition}
    &\mbbE\left[ \left(\widehat{\Cov}(\bar{Y}_{12}, \bar{Y}_{13}) -  \Cov(\bar{Y}_{12}, \bar{Y}_{13})\right)^2 \right] \nonumber \\
    &\quad= 
    \underbrace{\Var(\widehat{\Cov}(\bar{Y}_{12}, \bar{Y}_{13}))
    \vphantom{\left( \mbbE \left[ \widehat{\Cov}(\bar{Y}_{12}, \bar{Y}_{13}) \right] - \Cov(\bar{Y}_{12}, \bar{Y}_{13}) \right)^2}
    }_{\textbf{Term 3}}
    + 
    \underbrace{\left( \mbbE \left[ \widehat{\Cov}(\bar{Y}_{12}, \bar{Y}_{13}) \right] - \Cov(\bar{Y}_{12}, \bar{Y}_{13}) \right)^2}_{\textbf{Term 4}}.
\end{align}
To complete the proof, we will show below that that each of the two terms on the right hand side of \eqref{eq:cov_decomposition} are $O(G^{-1})$. 

\framebox{\textbf{Term 3}}
We start with the first term, the variance of the covariance estimator.  We can rewrite 
\begin{align*}\label{eq:covariance_est_cov}
    \Var(\widehat{\Cov}(\bar{Y}_{12}, \bar{Y}_{13})) = %
    \sum_{i=1}^G\sum_{\substack{j=1\\j\neq i}}^G\sum_{\substack{k=1\\k\neq i, j}}^G \sum_{l=1}^G\sum_{\substack{m=1\\m\neq l}}^G\sum_{\substack{n=1\\n\neq l, m}}^G\ddfrac{\Cov\left\{(\bar{Y}_{ij} - \hfmr)(\bar{Y}_{ik} - \hfmr), (\bar{Y}_{lm} - \hfmr) (\bar{Y}_{ln} - \hfmr))\right\}}{G^2(G-1)^2(G-2)^2}. 
\end{align*}
In order to show that it converges to $0$, we need to prove that the number of nonzero covariance terms is of the order smaller than $G^6$.

\begin{itemize}
    \item 
    Terms involving $\Cov(\bar{Y}_{ij}\bar{Y}_{ik}, \bar{Y}_{lm}\bar{Y}_{ln})$:
    These terms will be zero when all indices are different, that is in $G!/(G-6)!$ cases. Thus, $[G(G-1)(G-2)]^2 - G!/(G-6)!=O(G^5)$ of the terms in the sum above will be nonzero. 
    \item
    Terms involving $\Cov(\bar{Y}_{ij}\bar{Y}_{ik}, \bar{Y}_{lm}\hfmr)$:
    We have 
    \begin{align*}
        &\Cov(\bar{Y}_{ij}\bar{Y}_{ik}, \bar{Y}_{lm}\hfmr)  \\
        &\quad= \frac{1}{G(G-1)} \Cov\left\{\bar{Y}_{ij}\bar{Y}_{ik}, 2\bar{Y}_{lm}^2 + 4\sum_{\substack{n=1\\n\neq l,m}}^G\bar{Y}_{lm}\bar{Y}_{ln} + \sum_{\substack{n=1\\n\neq l,m}}^G\sum_{\substack{p=1\\ p\neq l,m,n}}^G \bar{Y}_{lm}\bar{Y}_{np}\right\}\\
        &\quad=  \frac{\Cov(\bar{Y}_{ij}\bar{Y}_{ik}, 2\bar{Y}_{lm}^2)}{G(G-1)} + 4\sum_{\substack{n=1\\n\neq l,m}}^G\frac{\Cov\left(\bar{Y}_{ij}\bar{Y}_{ik}, \bar{Y}_{lm}\bar{Y}_{ln}\right)}{G(G-1)} + \sum_{\substack{n=1\\n\neq l,m}}^G\sum_{\substack{p=1\\p\neq l,m,n}}^G \frac{\Cov(\bar{Y}_{ij}\bar{Y}_{ik}, \bar{Y}_{lm}\bar{Y}_{np})}{G(G-1)}. 
    \end{align*}
    The first term will be nonzero when $\bar{Y}_{ij}\bar{Y}_{ik}$ and $\bar{Y}_{lm}^2$ share any of the indices, hence 
    \begin{align*}
        &\frac{1}{G(G-1)}\sum_{i=1}^G\sum_{\substack{j=1\\j\neq i}}^G\sum_{\substack{k=1\\k\neq i, j}}^G \sum_{l=1}^G\sum_{\substack{m=1\\m\neq l}}^G\sum_{\substack{n=1\\n\neq l,m}}^G\Cov(\bar{Y}_{ij}\bar{Y}_{ik}, 2\bar{Y}_{lm}^2) \\
        &\quad = \frac{G(G-1)(G-2)^2}{G(G-1)}\sum_{l=1}^G \sum_{\substack{m=1\\m\neq l}}^G \Cov(\bar{Y}_{12}\bar{Y}_{13}, 2\bar{Y}_{lm}^2),
    \end{align*}
    which is $O(G^3)$. The second term is $O(G^4)$, while the third term is $O(G^5)$. %
    \item 
    Terms involving $\Cov(\hfmr^2, \hfmr^2)$: 
    We have 
    \begin{align*}
        &\Cov(\hfmr^2, \hfmr^2) \\
        &\quad = 
        \frac{1}{G^2(G-1)^2}
        \sum_{i=1}^G \sum_{\substack{j=1\\j \neq i}}^G 
        \Cov\left\{
        2\bar{Y}_{ij}^2 + 4\sum_{\substack{k=1\\k\neq i, j}}^G \bar{Y}_{ij}\bar{Y}_{ik} +\sum_{\substack{k=1\\k\neq i, j}}^G\sum_{\substack{l=1\\l\neq i, j,k}}^G \bar{Y}_{ij}\bar{Y}_{kl} , \hfmr^2 
        \right\} \\
        &\quad = \frac{1}{G(G-1)} \Cov\left\{(2\bar{Y}_{12}^2 + 4\sum_{\substack{k=1\\ k\neq 1, 2}}^G \bar{Y}_{12}\bar{Y}_{1k} + \sum_{\substack{k=1\\ k\neq 1,2}}^G \sum_{\substack{l=1\\l\neq 1,2,k}}^G \bar{Y}_{12}\bar{Y}_{kl}, \hfmr^2\right\}\\
        &\quad = \frac{1}{G(G-1)} \Cov\left\{(2\bar{Y}_{12}^2 + 4(G-2) \bar{Y}_{12}\bar{Y}_{13} + (G-2)(G-3)\bar{Y}_{12}\bar{Y}_{34}, \hfmr^2\right\}. 
    \end{align*}
    The leading term in this expression is
    \begin{gather*}
    \frac{(G-2)(G-3)}{G^3(G-1)^3}\sum_{i=1}^G\sum_{\substack{j=1\\j\neq i}}^G\sum_{\substack{k=1\\k\neq i, j}}^G\sum_{\substack{l=1\\l\neq i, j, k}}^G \Cov(\bar{Y}_{12}\bar{Y}_{34}, \bar{Y}_{ij}\bar{Y}_{kl}) = O(G^{-1}). 
    \end{gather*}  
        \item
    Terms involving $\Cov(\bar{Y}_{ij}\bar{Y}_{ik}, \hfmr^2)$ and $\Cov(\bar{Y}_{ij}\hfmr, \hfmr^2)$: 
    These terms are handled in a similar manner and their proofs are omitted. 
\end{itemize}

Thus, we have thus shown that 
\begin{gather}\label{eq:term1_order}
     \Var(\widehat{\Cov}(\bar{Y}_{12}, \bar{Y}_{13})) = O(G^5/G^6) = O(G^{-1}). 
\end{gather}

\framebox{\textbf{Term 4}}
We now turn to the second term, which is the squared bias. We have 
\begin{gather*}
    \mbbE 
    \left[
    \widehat{\Cov}(\bar{Y}_{12}, \bar{Y}_{13})) 
    \right]
    =  \left[ 1 - \frac{4(G-2)}{G(G-1)}\right] \mbbE[\bar{Y}_{12}\bar{Y}_{13}] - \frac{2}{G(G-1)}\mbbE[\bar{Y}_{12}^2] - \fmr^2 \frac{(G-2)(G-3)}{G(G-1)}. 
\end{gather*}
It follows that the bias is given by 
\begin{gather*}
   \mbbE 
   \left[
   \widehat{\Cov}(\bar{Y}_{12}, \bar{Y}_{13})
   \right]
   - \Cov (\bar{Y}_{12}, \bar{Y}_{13}) =  - \frac{4(G-2)}{G(G-1)} \mbbE[\bar{Y}_{12}\bar{Y}_{13}] - \frac{2}{G(G-1)}\mbbE[\bar{Y}_{12}^2]  - \frac{2(2G - 3)}{G(G-1)}\fmr^2. 
\end{gather*}
Thus, we have
\begin{gather}\label{eq:term2_order}
    \left( \mbbE \widehat{\Cov}(\bar{Y}_{12}, \bar{Y}_{13}) - \Cov (\bar{Y}_{12}, \bar{Y}_{13})\right)^2 = O(G^{-2}),
\end{gather}
which goes to $0$ as $G\rightarrow\infty$. 

Putting \eqref{eq:term1_order} and \eqref{eq:term2_order} together,
along with \eqref{eq:chebychev_cov},
the result then follows.
This completes the proof of the consistency of
$\widehat{\Cov}(\bar{Y}_{12}, \bar{Y}_{13})$.

\subsubsection{Proof of \Cref{prop:equivalence_jackk_pi}
(equivalence of plug-in and jackknife variance estimators)}
Recall the form of the jackknife variance estimator
of $\Var(\sqrt{G} \hfmr)$:
\begin{equation}
    \label{eq:hvarhfrr-jackknife}
     \widehat{\Var}_{JK}(\sqrt{G}\hfmr) = \frac{(G-2)^2}{G} \sum_{i=1}^G (\hfmr_{-i} - \hfmr)^2 - 2\frac{\widehat{\Var}(\bar{Y}_{12})}{G-1},
\end{equation}
where $\hfmr_{-i}$ is defined as
\begin{equation}
    \label{eq:hfrr-loo}
    \hfmr_{-i} 
    = 
    \frac{1}{(G - 1) (G - 2)}
    \sum_{j=1}^G \sum_{\substack{k=1,\\k\neq j}}^{G} \bar{Y}_{jk}\mathds{1}(\{j\neq i\}\cap  \{k\neq i\}).
\end{equation}
Recall also the estimator for $\Var(\bar{Y}_{12})$:
\begin{equation}
    \label{eq:hvarbary12}
    \widehat{\Var}(\bar{Y}_{12}) = \frac{1}{G(G-1)}\sum_{i=1}^G\sum_{\substack{j=1,\\j\neq i}}^{G} (\bar{Y}_{ij}-\hfmr)^2.
\end{equation}
Through a series of algebraic manipulations,
we will show that 
after substituting for \eqref{eq:hfrr-loo} and \eqref{eq:hvarbary12},
the expression \eqref{eq:hvarhfrr-jackknife}
simplifies to plug-in estimator from \eqref{eq:varest_plugin}.

Towards that end,
we start by expanding the sum in the first term
on the right-hand side of \eqref{eq:hvarhfrr-jackknife}:
\begin{align}
 &\sum_{i=1}^G ( \hfmr_{-i} - \hfmr)^2 \nonumber \\
 &\quad =  \sum_{i=1}^G \left( \frac{\sum_{k=1}^G\sum_{\substack{l=1\\l \neq k}}^G \bar{Y}_{kz} - 2\sum_{\substack{j=1\\j\neq i}}^G \bar{Y}_{ij}}{(G-1)(G-2)} - \hfmr \right)^2 \nonumber \\
 &\quad = \frac{4}{(G-2)^2} \sum_{i=1}^G \left(\sum_{\substack{j=1\\j\neq i}}^G\frac{\bar{Y}_{ij}}{G-1} - \hfmr \right)^2 \nonumber \\
&\quad = \frac{4}{(G-2)^2(G-1)^2}\sum_{i=1}^G\left[\sum_{\substack{j=1\\j\neq i}}^G(\bar{Y}_{ij} - \hfmr)^2 +\sum_{\substack{j=1\\j\neq i}}^G\sum_{\substack{k=1\\k\neq i, j}}^G(\bar{Y}_{ij} - \hfmr)(\bar{Y}_{ik} - \hfmr) \right]. \label{...}
\end{align}

Moving the appropriate factor across
and subtracting the second term
on the right-hand side of \eqref{eq:hvarhfrr-jackknife},
we arrive at
\begin{align}
    &\frac{(G-2)^2}{G} \sum_{i=1}^G (\hfmr_{-i} - \hfmr)^2 
    - 2\frac{\widehat{\Var}(\bar{Y}_{12})}{G-1}
    \nonumber 
    \\ &\quad = \frac{4\sum_{i=1}^G\sum_{\substack{j=1\\j\neq i}}^G(\bar{Y}_{ij} - \hfmr)^2}{G(G-1)^2} + \frac{4\sum_{i=1}^G\sum_{\substack{j = 1\\j\neq i}}^G\sum_{\substack{k = 1\\k\neq i, j}}^G(\bar{Y}_{ij}-\hfmr)(\bar{Y}_{ik} - \hfmr)}{G(G-1)^2}
    - 2\frac{\widehat{\Var}(\bar{Y}_{12})}{G-1} \nonumber \\
    &\quad = \frac{2\widehat{\Var}(\bar{Y}_{12}) +  4\widehat{\Cov}(\bar{Y}_{12}, \bar{Y}_{13})(G-2)}{G-1} + 2\frac{\widehat{\Var}(\bar{Y}_{12})}{G-1}
    - 2\frac{\widehat{\Var}(\bar{Y}_{12})}{G-1} \nonumber \\
    &\quad =
     \frac{2}{G-1}
      \widehat{\Var}(\bar{Y}_{12})
      +
      \frac{4 (G-2)}{G-1} \widehat{\Cov}(\bar{Y}_{12}, \bar{Y}_{13}), 
      \label{eq:varest-plug-pf}
\end{align}
Noting that the \eqref{eq:varest-plug-pf}
matches with \eqref{eq:varest_plugin}.
we have that $ \widehat{\Var}_{JK}(\sqrt{G} \hfmr) = \widehat{\Var}(\sqrt{G} \hfmr)$,
as claimed.

\subsection{Proofs for resampling-based methods in \Cref{sec:methods_nonparametric}}
\label{sec:proofs_nonparametric}

\subsubsection{Proof of \Cref{prop:bootstrap_var_subsets}
(bias of subsets bootstrap estimators)}
Recall that $\hfnmr_b^*$ and $\hfmr_b^*$ indicate the $\fnmr$ and $\fmr$ estimates respectively based on the $b$-th bootstrap sample. 
The proofs for various statements in the proposition are separated below.

\begin{itemize}
    \item 
    Showing that $\Bias(\bhfnmr) = 0$ and $\Bias(\Var^*(\sqrt{G}\hfnmr^*)) = - \Var(\hfnmr)$ is straightforward. 
    \begin{itemize}
        \item 
        For $\Bias(\bhfnmr)$, it is easy to see that 
        \begin{align*}
            \mbbE[\mbbE^*[\bhfnmr]] = \frac{1}{G}\sum_{i=1}^G\mbbE[\mbbE^*[W_i]\bar{Y}_{ii}] = \frac{1}{G}\mbbE[\bar{Y}_{ii}] = \fnmr. 
        \end{align*}
        Hence,
        $\Bias(\bhfnmr) = 0$.
        \item
        Towards computing $\Bias(\sqrt{G} \bVar(\bhfnmr))$,
        observe that
        \begin{align*}
            \mbbE[\bVar[\bhfnmr]]
            &= \frac{1}{G^2}\sum_{i=1}^G \mbbE\left\{\bVar(W_i)\bar{Y}_{ii}^2 + \bCov(\bar{W}_{i}, \bar{W}_{k})\sum_{\substack{k=1\\k\neq i}}^G\bar{Y}_{ii}\bar{Y}_{kk}\right\} \\
            &= \frac{1}{G^2}\sum_{i=1}^G \mbbE\left\{ \frac{G-1}{G} \bar{Y}_{ii}^2 - \frac{1}{G}\sum_{\substack{k=1\\k\neq i}}^G\bar{Y}_{ii}\bar{Y}_{kk}\right\}\\
            &= \frac{G-1}{G^2}\mbbE[\bar{Y}_{11}^2] - \frac{G-1}{G^2}\fnmr \\
            &= \frac{G-1}{G}\Var(\hfnmr). 
        \end{align*}
        Thus, we have
        $
            \Bias(\Var^*(\sqrt{G} \hfnmr))
            = (G - 1) \Var(\hfnmr)
            - G \Var(\hfnmr)
            = - \Var(\hfmr),
        $
        as claimed.
    \end{itemize}
    \item
    Obtaining expressions for $\Bias(\bhfmr)$ and $\Bias(\Var^*(\sqrt{G} \bhfmr))$
    is slightly more involved.
    \begin{itemize}
        \item
        For $\Bias(\bhfmr)$, note that 
        \begin{align*}
            \mbbE[\mbbE^*[\bhfmr]] =\frac{1}{G(G-1)}\sum_{i=1}^G\mbbE\left[\mbbE^*[W_i] \sum_{\substack{j=1\\j\neq i}}^G \bar{Y}_{ij} \right] = \frac{1}{G(G-1)}\sum_{i=1}^G\sum_{\substack{j=1\\j\neq i}}^G\mbbE\left[ \bar{Y}_{ij}\right] = \fmr. 
        \end{align*}
        Hence,
        $\Bias(\bhfmr) = 0$.
        \item
        For $\bVar(\bhfmr)$, observe that
        \begin{align*}
        &\mbbE[\bVar[\bhfmr]] \\
        &\quad = 
         \frac{1}{G^2}\sum_{i=1}^G \mbbE\left\{ \frac{\left(\sum_{\substack{j=1\\j \neq i}}^G  \bar{Y}_{ij}\right)^2}{(G-1)^2} \bVar\left( W_i \right)  + 
         \sum_{\substack{k=1\\k\neq i}}^G \frac{\sum_{\substack{j=1\\j \neq i}}^G  \bar{Y}_{ij}\sum_{\substack{l=1\\l \neq k}}^G  \bar{Y}_{kl}}{(G-1)^2} \bCov\left( W_i, W_k \right) \right\}\\
        &\quad =  \frac{1}{G^2}\sum_{i=1}^G \mbbE\left\{ \frac{\left(\sum_{\substack{j=1\\j \neq i}}^G  \bar{Y}_{ij}\right)^2}{(G-1)^2} \frac{G-1}{G}  - 
         \sum_{\substack{k=1\\k\neq i}}^G \frac{\sum_{\substack{j=1\\j \neq i}}^G  \bar{Y}_{ij}\sum_{\substack{l=1\\l \neq k}}^G  \bar{Y}_{kl}}{(G-1)^2} \frac{1}{G}  \right\}. 
         \end{align*}
         We thus have
         \begin{align}
         &\mbbE[\bVar[\bhfmr]] \nonumber \\
         &\quad=
         \frac{1}{G^2}\sum_{i=1}^G \mbbE\left\{ \frac{\left(\sum_{\substack{j=1\\j \neq i}}^G  \bar{Y}_{ij}\right)^2}{(G-1)^2} - \hfmr^2 \right\}  \nonumber\\
        &\quad = 
        \frac{1}{G} \mbbE\left\{\frac{\bar{Y}_{12}^2 + (G-2)\bar{Y}_{12}\bar{Y}_{13}}{G-1} - \hfmr^2\right\} \nonumber \\
        &\quad =
        \frac{1}{G} \mbbE\Bigg\{\frac{\bar{Y}_{12}^2 + (G-2)\bar{Y}_{12}\bar{Y}_{13}}{G-1} \nonumber \\
        &\qquad \qquad \qquad - \left[\frac{2\bar{Y}_{12}^2 + 4(G-2)\bar{Y}_{12}\bar{Y}_{13} + (G-2)(G-3)\bar{Y}_{12}\bar{Y}_{34}}{G(G-1)} \right]\Bigg\}.  \label{eq:prop4_finalterm}
        \end{align}
        We can rewrite the first of the two terms in \eqref{eq:prop4_finalterm} as
        \begin{align}
            &\frac{1}{G} \mbbE\left\{\frac{\bar{Y}_{12}^2 + (G-2)\bar{Y}_{12}\bar{Y}_{13}}{G-1}\right\} \nonumber \\ 
            &\quad= \Var(\hfmr) - \frac{\Var(\bar{Y}_{12}) + 3(G-2)\Cov(\bar{Y}_{12}, \bar{Y}_{13})}{G(G-1)} +  \frac{\fmr^2}{G}, \label{eq:prop4_ft}
        \end{align}
        and the second as 
        \begin{gather}\label{eq:prop4_st}
            \frac{1}{G}\mbbE\left\{ \frac{2\bar{Y}_{12}^2 + 4(G-2)\bar{Y}_{12}\bar{Y}_{13} + (G-2)(G-3)\bar{Y}_{12}\bar{Y}_{34}}{G(G-1)} \right\} = \frac{\Var(\hfmr)}{G}  +\frac{\fmr^2}{G}. 
        \end{gather}
        Thus, combining  \eqref{eq:prop4_ft} and \eqref{eq:prop4_st} with \eqref{eq:prop4_finalterm}, we obtain 
        \begin{align*}
        \mbbE[\bVar[\bhfmr]]
        = 
        \frac{G-1}{G}\Var(\hfmr) - \frac{\Var(\bar{Y}_{12}) + 3(G-2)\Cov(\bar{Y}_{12}, \bar{Y}_{13})}{G(G-1)}. %
        \end{align*}
        Therefore,
        we have
        \begin{align*}
            \Bias(\Var^*(\sqrt{G} \bhfnmr))
            &=
            (G - 1) \Var(\hfmr) - \frac{\Var(\bar{Y}_{12}) + 3(G-2)\Cov(\bar{Y}_{12}, \bar{Y}_{13})}{(G-1)}
            - G \Var((\hfmr) \\
            &= 
            - \Var(\hfmr) - \frac{\Var(\bar{Y}_{12}) + 3(G-2)\Cov(\bar{Y}_{12}, \bar{Y}_{13})}{(G-1)},
        \end{align*}
        as promised.
    \end{itemize}
    \end{itemize}

This completes the bias derivations for subsets bootstrap estimators.

\subsubsection{Proof of \Cref{prop:bootstrap_var_vertex}
(bias of vertex bootstrap estimators)}

Recall from \eqref{eq:hfrr-vertex}
the expression for $\bhfmr$,
the estimator for $\fmr$ based on the $b$-th bootstrap sample using vertex bootstrap:
\begin{align*}
    \bhfmr 
    = \sum_{i, j=1}^G W_i\bigg[ \frac{(W_i - 1) \hfmr \mathds{1}(i=j)}{G(G-1)}
    + \frac{W_j \bar{Y}_{ij} \mathds{1}(i\neq j) }{G(G-1)}\bigg].
\end{align*}

\begin{itemize}
\item We start with deriving $\Bias(\bhfmr)$. Note that 
\begin{align*}
    &\mbbE\left\{ \mbbE^*[\bhfmr] \right\} \\
    &\quad = \frac{1}{G(G-1)}\sum_{i, j=1}^G\mbbE\left\{ \mbbE^*\left[ W_i(W_i-1)\mathds{1}(i=j) \right] \hfmr + \bar{Y}_{ij} \mbbE^*\left[W_iW_j \mathds{1}(i \neq j) \right] \right\}\\
    &\quad = \frac{1}{G(G-1)}\sum_{i=1}^G\mbbE^*\left[ \hfmr  \frac{G-1}{G} + \sum_{\substack{j=1\\j \neq i}}^G\bar{Y}_{ij} \frac{G-1}{G} \right]\\
    &\quad = \frac{1}{G(G-1)} \left[ G\fmr\frac{G-1}{G} + G(G-1)\fmr \frac{G-1}{G}\right]\\
    &\quad = \fmr. 
\end{align*}
Thus, $\Bias(\bhfmr) = 0$.
\item We next turn to deriving $\Bias(\bVar(\bhfmr))$. 
Let $\bar{Y}^*_{ij}$ denote the observation corresponding to the $i$-th and $j$-th identities in the $b$-th bootstrap sample, where the subscript $b$ is omitted.
It is easy to see that 
\begin{align*}
    &\mbbE\left\{\bVar(\bhfmr)\right\} \\
    &\quad= \frac{1}{G(G-1)}\mbbE\bigg\{2 \bVar(\bar{Y}_{12}^*) + 4(G-2)\bCov\left(\bar{Y}_{12}^*, \bar{Y}_{13}^*\right) + (G-2)(G-3)\bCov\left(\bar{Y}_{12}^*, \bar{Y}_{34}^*\right)\bigg\}. 
\end{align*}
For the variance term $\bVar(\bar{Y}_{12}^*)$, we have 
\begin{align}
    \mbbE\left\{\bVar(\bar{Y}_{12}^*)\right\}
    & = \mbbE\left\{\mbbE^*\left[\bar{Y}_{12}^{2*}\right] - \mbbE^*\left[\bar{Y}_{12}^*\right]^2\right\}\nonumber \\
    & = \frac{G-1}{G}\mbbE\left[\bar{Y}_{12}^2\right] + \frac{1}{G}\mbbE\left[\hfmr^2\right] - \mbbE\left[\hfmr^2\right] \nonumber 
    \\
    & = \frac{G-1}{G}\left\{\mbbE[\bar{Y}_{12}^2] - \mbbE[\hfmr^2] \right\} \nonumber 
    \\ & = \frac{G-1}{G} \left\{\Var(\bar{Y}_{12})  - \Var(\hfmr)  \right\} \label{eq:varvertex_y12}.     %
\end{align}
For the covariance term $\bCov\left(\bar{Y}_{12}^*, \bar{Y}_{34}^*\right)$, we can show that  
\begin{align*}
   \mbbE\left\{\bCov(\bar{Y}_{12}^*, \bar{Y}_{13}^*)\right\}
   &= \mbbE\left\{\mbbE^*[\bar{Y}_{12}^*\bar{Y}_{13}^*] - \mbbE^*[\bar{Y}_{12}^*]\mbbE^*[\bar{Y}_{13}^*]\right\}\\
   & = \mbbE\left\{ \frac{2G-1}{G^2}\hfmr^2 + \frac{(G-1)^2}{G^2} \left( \frac{1}{G-1}\bar{Y}_{12}^2 + \frac{G-2}{G-1}\bar{Y}_{12}\bar{Y}_{13}\right) - \hfmr^2\right\}\\
   & = \mbbE\left\{ \frac{(G-1)^2}{G^2} \left( \frac{1}{G-1}\bar{Y}_{12}^2 + \frac{G-2}{G-1}\bar{Y}_{12}\bar{Y}_{13}\right) - \frac{(G-1)^2}{G^2} \hfmr^2\right\} \\
    &=\frac{(G-1)^2}{G}\frac{1}{G}\mbbE\left\{  \frac{\bar{Y}_{12}^2}{G-1} + \frac{(G-2)\bar{Y}_{12}\bar{Y}_{13}}{G-1} - \hfmr^2\right\}. 
\end{align*}
By following the same derivation as in \eqref{eq:prop4_finalterm}, we can further show that 
\begin{align}\label{eq:covvertex_y12_y13}
& \mbbE\left\{\bCov(\bar{Y}_{12}^*, \bar{Y}_{13}^*)\right\} 
= \frac{(G-1)^2}{G} \left\{ \frac{G-1}{G}\Var(\hfmr) - \frac{\Var(\bar{Y}_{12}) + 3(G-2)\Cov(\bar{Y}_{12}, \bar{Y}_{13})}{G(G-1)}\right\}.
\end{align}

Thus, combining \eqref{eq:varvertex_y12} and \eqref{eq:covvertex_y12_y13}, together with the fact that $\Cov(\bar{Y}_{12}^*, \bar{Y}_{34}^*) = 0$ (by independence), yields
\begin{align*}
    &\mbbE\left\{\bVar(\bhfmr)\right\}\\
    &\quad = \frac{1}{G(G-1)} \bigg\{  2\frac{G-1}{G}\left[\Var(\bar{Y}_{12}) - \Var(\hfmr) \right] \\
    &\quad \quad + \frac{4(G-1)^2(G-2)}{G}\left[  \frac{G-1}{G}\Var(\hfmr) - \frac{\Var(\bar{Y}_{12}) + 3(G-2)\Cov(\bar{Y}_{12}, \bar{Y}_{13})}{G(G-1)} \right]\bigg\}\\
    &\quad =  \frac{2}{G^2}\left[ \Var(\bar{Y}_{12}) - \Var(\hfmr)\right] \\
    &\quad \quad + \frac{4(G-1)(G-2)}{G^2}\left[  \frac{G-1}{G}\Var(\hfmr) - \frac{\Var(\bar{Y}_{12}) + 3(G-2)\Cov(\bar{Y}_{12}, \bar{Y}_{13})}{G(G-1)} \right]\\
    &\quad = \frac{2}{G^2}\left[ \Var(\bar{Y}_{12}) - \Var(\hfmr)\right]\\ 
    &\quad\quad + \frac{4(G-1)(G-2)}{G^2} \left[ \frac{\Var(\bar{Y}_{12}) + (G-2)\Cov(\bar{Y}_{12}, \bar{Y}_{13})}{G(G-1)} - \frac{\Var(\hfmr)}{G} \right]
    \\
     &\quad = \Var(\hfmr) - \frac{2\Var(\bar{Y}_{12})}{G^2(G-1)} - \frac{2\Var(\hfmr)}{G^2} \\ 
    &\quad\quad - \frac{4(3G-2)(G-2)}{G^3(G-1)}\Cov(\bar{Y}_{12}, \bar{Y}_{13}) + \frac{4(G-2)}{G^3}\Var(\bar{Y}_{12}) - \frac{4(G-1)(G-2)}{G^3}\Var(\hfmr).
\end{align*}
We can rearrange the terms to obtain 
\begin{align}
\begin{aligned}
 &\Bias(\bVar(\bhfmr)\\
    &\quad = \Var(\bar{Y}_{12}) \left[\frac{4(G-2)}{G^3} + O\left( G^{-3}\right) \right] + \Cov(\bar{Y}_{12}, \bar{Y}_{13})\left[ 
-\frac{28}{G(G-1)} + O(G^{-3}) \right] \label{eq:bias-var-vertex-frr}.
\end{aligned}
\end{align}
Up to constants, the expression in \eqref{eq:bias-var-vertex-frr} matches with the expression in the statement.
\end{itemize}
This completes the derivations of the bias for the vertex bootstrap estimators.

\subsubsection{Proof of \Cref{prop:bootstrap_var_doublenothing}
(bias of double-or-nothing bootstrap estimators)}
Recall from \eqref{eq:db_estimators} the expression for $\bhfnmr$ and $\bhfmr$,
the estimator for $\fnmr$ and $\fmr$ based on the $b$-th bootstrap sample using double-or-nothing bootstrap:
\begin{align*}
    \hfnmr_b^* = \frac{\sum_{i=1}^G W_i \bar{Y}_{ii}}{\sum_{i=1}^G W_i},
    \quad 
    \text{and}
    \quad
    \bhfmr = \frac{\sum_{\substack{i, j=1\\j\neq i}}^G W_iW_j\bar{Y}_{ij}}{\sum_{\substack{i, j=1\\j\neq i}}^G W_iW_j},
\end{align*}
where $\mbbE[W_i] = 1$, $\Var(W_i) = \tau$, and $W_i$ is independent of $W_j$ whenever $i\neq j$ for $i, j\in \cG$. The double-or-nothing bootstrap falls in this framework when $\tau = 1$. 

\begin{itemize}
\item It it straightforward to show that $\Bias(\bhfnmr) = 0$. Next, we examine $\Bias(\bVar(\bhfnmr))$. Let $T^*=\sum_{i=1}^G W_i\bar{Y}_{ii}$ and $N^*=\sum_{i=1}^GW_i$. Through an application of the Delta method, we obtain 
\begin{align*}
    \mbbE\left\{\bVar(\bhfnmr)\right\} =  \frac{1}{G^2}\mbbE\left\{\bVar(T^*) - 2\hfnmr \bCov(T^*, N^*) + \hfnmr^2 \bVar(N^*)\right\},
\end{align*}
where
\begin{align*}
    \mbbE[\bVar(T^*)]
    &= \mbbE\left[\sum_{i=1}^G\bar{Y}_{ii}^2\tau\right] = \tau G\mbbE[\bar{Y}_{11}^2], \\
    \mbbE[\hfnmr\bCov(T^*, N^*)] 
    &=  \mbbE[\hfnmr^2\bVar(N^*)] = G\tau\mbbE[\hfnmr^2] = \tau \left[\mbbE[\bar{Y}_{11}^2] + \fnmr^2(G-1) \right]. 
\end{align*}
Hence,
we have
\begin{gather*}
    \mbbE[\bVar(\bhfnmr)] =  \frac{G-1}{G} \tau \Var(\hfnmr). 
\end{gather*}
Taking $\tau=1$ yields the result. 

\item With respect to the $\fmr$, let $T^*=\sum_{i=1}^G\sum_{\substack{j=1\\j\neq i}}^GW_i W_j\bar{Y}_{ij}$ and $N^*=\sum_{i=1}^G\sum_{\substack{j=1\\j\neq i}}^GW_iW_j$. Again, it is easy to see that $\Bias(\bhfmr) = 0$. An application of the Delta method yields 
\begin{gather*}
    \mbbE\left\{\bVar(\bhfmr)\right\} = \frac{1}{G^2(G-1)^2}\mbbE\left\{\bVar(T^*) - 2\hfmr\bCov(T^*, N^*) + \hfmr^2 \bVar(N^*)\right\},
\end{gather*}
where
\begin{align*}
     \mbbE[\bVar(T^*)]
     &=G(G-1)\mbbE\left\{ 2\bVar(\bar{Y}_{12}W_1W_2) + 4(G-2)\bCov(\bar{Y}_{12}W_1W_2, \bar{Y}_{13}W_1W_3) \right\}\\
     &=G(G-1)\mbbE\left\{ 2\bar{Y}_{12}(\mbbE^*[W_1^2]\mbbE^*[W_2^2]-1) + 4(G-2)\bar{Y}_{12}\bar{Y}_{13}(\mbbE^*[W_1^2] - 1)\right\}\\
     &=G(G-1) [2\tau(\tau + 2) \mbbE[\bar{Y}_{12}^2] + 4(G-2)\tau\mbbE[\bar{Y}_{12}\bar{Y}_{13}]],
\end{align*}
and 
\begin{align*}
     \mbbE\left[\hfmr \bCov(T^*, N^*)\right]  = 
     \mbbE[\hfmr^2 \bVar(N^*)] = G(G-1)[2\tau (\tau + 2) + 4(G-2)\tau ]\mbbE[\hfmr^2]. 
\end{align*}
It then follows that 
\begin{align*}
    \mbbE[\bVar(\bhfmr)]
    &= \frac{1}{G(G-1)}\left[ 2\tau (\tau + 2)\mbbE[\bar{Y}_{12}^2 - \hfmr^2] + 4(G-2)\tau \mbbE[\bar{Y}_{12}\bar{Y}_{13} - \hfmr^2] \right]\\
    &= \frac{1}{G(G-1)}\left[ 2\tau (\tau + 2)\Var(\bar{Y}_{12}) + 4(G-2)\tau \Cov(\bar{Y}_{12}, \bar{Y}_{13})\right] \\ &\quad\quad - \frac{2\tau (\tau + 2G - 2)}{G(G-1)}\frac{2\Var(\bar{Y}_{12}) + 4(G-2)\Cov(\bar{Y}_{12}, \bar{Y}_{13})}{G(G-1)}.
\end{align*}
Choosing $\tau = 1$ yields
\begin{align*}
\Bias(\bVar(\bhfmr)) = -\Var(\hfmr) \frac{2(2G -1)}{G(G-1)} + \frac{4\Var(\bar{Y}_{12})}{G(G-1)}. 
\end{align*}
\end{itemize}
This completes the bias derivations for the double-or-nothing bootstrap estimators.

\subsection{Proofs for unbalanced setting in \Cref{sec:unbalanced_setting}}
\label{sec:proofs_unbalanced}

\subsubsection[Proof of \Cref{prop:unbalanced_var_boot}]{Proof of \Cref{prop:unbalanced_var_boot}
(consistency of bootstrap estimators for $\fnmr$)} 

We separate the proof into consistency of 
subsets and vertex bootstrap,
and that of double-or-nothing bootstrap below.

\begin{itemize}
    \item \textbf{Consistency of subsets and vertex bootstrap estimators.} The resampling performed by these two bootstrap types for $\fnmr$ computations is analogous, thus we investigate consistency of both types altogether. By applying the Delta method, we obtain 
\begin{align*}
    &\bVar(\bhfnmr) \\
    &\quad= \frac{\sum_{i=1}^G \tilde{M}_i^2 \bar{Y}_{ii}^2 - \left(\frac{1}{\sqrt{G}}\sum_{i=1}^G \tilde{M}_i\bar{Y}_{ii}\right)^2}{\left(\sum_{i=1}^G \tilde{M}_i\right)^2} \\
    &\quad\quad - 2 \frac{ \left(\sum_{i=1}^G \tilde{M}_i \bar{Y}_{ii}\right) \left[ \sum_{i=1}^G \tilde{M}_i^2 \bar{Y}_{ii} - (\sum_{i=1}^G \tilde{M}_i \bar{Y}_{ii}) (\sum_{i=1}^G \tilde{M}_i/G)\right]}{\left(\sum_{i=1}^G \tilde{M}_i\right)^3}\\
   &\quad\quad\quad + \frac{\left(\sum_{i=1}^G \tilde{M}_i \bar{Y}_{ii}\right)^2 \left[\sum_{i=1}^G \tilde{M}_i^2 - \left(\frac{1}{\sqrt{G}}\sum_{i=1}^G \tilde{M}_i\right)^2\right]}{\left(\sum_{i=1}^G \tilde{M}_i\right)^4}. 
\end{align*}
Since $M_i$ is finite for any $i\in\cG$, we can apply the weak law of large numbers and the continuous mapping theorem to obtain 
the following convergences in probability as $G \to \infty$:
\begin{align*}
      G\frac{\sum_{i=1}^G \tilde{M}_i^2 \bar{Y}_{ii}^2 - \left(\frac{1}{\sqrt{G}}\sum_{i=1}^G \tilde{M}_i\bar{Y}_{ii}\right)^2}{\left(\sum_{i=1}^G \tilde{M}_i\right)^2}  &\pto  \frac{\Var(\tilde{M}_1\bar{Y}_{11})}{\mbbE[\tilde{M}_1]^2}, \\
      G\frac{\left(\sum_{i=1}^G \tilde{M}_i \bar{Y}_{ii}\right)\sum_{i=1}^G \left( \tilde{M}_i^2 \bar{Y}_{ii} - (\sum_{i=1}^G \tilde{M}_i \bar{Y}_{ii}) (\sum_{i=1}^G \tilde{M}_i/G)\right)}{\left(\sum_{i=1}^G \tilde{M}_i\right)^3} &\pto  \frac{\mbbE[\tilde{M}_1\bar{Y}_{11}] \Cov(\tilde{M}_1\bar{Y}_{11}, \tilde{M}_1)}{\mbbE[\tilde{M}_1]^3}, \\
      G\frac{\left(\sum_{i=1}^G \tilde{M}_i \bar{Y}_{ii}\right)^2 \left[\sum_{i=1}^G \tilde{M}_i^2 - \left(\frac{1}{\sqrt{G}}\sum_{i=1}^G \tilde{M}_i\right)^2\right]}{\left(\sum_{i=1}^G \tilde{M}_i\right)^4} &\pto \frac{\mbbE[\tilde{M}_1 \bar{Y}_{11}]^2\Var(\tilde{M}_1)}{\mbbE[\tilde{M}_1]^4}.
\end{align*}
It then follows that
\begin{gather*}
   \bVar(\sqrt{G}\bhfnmr)\pto \frac{\Var(\tilde{M}_1\bar{Y}_{11})}{\mbbE[\tilde{M}_1]^2} - 2\frac{\mbbE[\tilde{M}_1\bar{Y}_{11}] \Cov(\tilde{M}_1\bar{Y}_{11}, \tilde{M}_1)}{\mbbE[\tilde{M}_1]^3} + \frac{\mbbE[\tilde{M}_1 \bar{Y}_{11}]^2\Var(\tilde{M}_1)}{\mbbE[\tilde{M}_1]^4}. 
\end{gather*}
This completes the proof for the consistency of the subsets and vertex bootstrap estimators. 

\item  \textbf{Consistency of double-or-nothing bootstrap estimator.} Assume that $\mbbE[W_i] = 1$ and $\Var(W_i) = \tau$. In addition, let $W_i\ind W_j$ whenever $i\neq j$ for $i, j\in\cG$. The double-or-nothing bootstrap is obtained by taking $\tau = 1$. By applying the Delta method, we obtain 
\begin{gather*}
    \bVar(\bhfnmr) = \tau \frac{\sum_{i=1}^G \tilde{M}_i^2\bar{Y}_{ii}^2}{\left(\sum_{i=1}^G\tilde{M}_i\right)^2} - 2\tau \frac{\left( \sum_{i=1}^G \tilde{M}_i^2\bar{Y}_{ii}\right)\left( \sum_{i=1}^G \tilde{M}_i \bar{Y}_{ii} \right)}{\left(\sum_{i=1}^G\tilde{M}_i\right)^3} + \tau \frac{\left( \sum_{i=1}^G \tilde{M}_i^2\right)\left(\sum_{i=1}^G \tilde{M}_i \bar{Y}_{ii} \right)^2 }{\left(\sum_{i=1}^G\tilde{M}_i\right)^4}. 
\end{gather*}
Since $\tilde{M}_i$ is finite, we can apply the weak law of large numbers and the continuous mapping theorem to obtain, as $G\rightarrow\infty$, 
\begin{align*}
    \tau G \frac{\sum_{i=1}^G \tilde{M}_i^2\bar{Y}_{ii}^2}{\left(\sum_{i=1}^G\tilde{M}_i\right)^2}
    &\pto \tau \frac{\mbbE[\tilde{M}_1^2\bar{Y}_{11}^2]}{\mbbE[\tilde{M}_1]^2},  \\   
    2G\tau \frac{\left( \sum_{i=1}^G \tilde{M}_i^2\bar{Y}_{ii}\right)\left( \sum_{i=1}^G \tilde{M}_i \bar{Y}_{ii} \right)}{\left(\sum_{i=1}^G\tilde{M}_i\right)^3}
    &\pto 2\tau \frac{\mbbE[\tilde{M}_1^2\bar{Y}_{11}]\mbbE[\tilde{M}_1\bar{Y}_{11}]}{\mbbE[\tilde{M}_1]^3}, \\ %
    \tau G \frac{\left( \sum_{i=1}^G \tilde{M}_i^2\right)\left(\sum_{i=1}^G \tilde{M}_i \bar{Y}_{ii} \right)^2 }{\left(\sum_{i=1}^G\tilde{M}_i\right)^4}
    &\pto \tau \frac{\mbbE[\tilde{M}_1^2]\mbbE[\tilde{M}_1\bar{Y}_{11}]^2}{\mbbE[\tilde{M}_1]^4}.
\end{align*}   
Putting everything together, 
as $G \to \infty$,
we have that
\begin{align*}
     \bVar(\sqrt{G}\bhfnmr)
     &\pto\tau \frac{\mbbE[\tilde{M}_1^2\bar{Y}_{11}^2]}{\mbbE[\tilde{M}_1]^2} - 2\tau \frac{\mbbE[\tilde{M}_1^2\bar{Y}_{11}]\mbbE[\tilde{M}_1\bar{Y}_{11}]}{\mbbE[\tilde{M}_1]^3} + \tau \frac{\mbbE[\tilde{M}_1^2]\mbbE[\tilde{M}_1\bar{Y}_{11}]^2}{\mbbE[\tilde{M}_1]^4}\\
     &= \tau \frac{\Var(\tilde{M}_1\bar{Y}_{11})}{\mbbE[\tilde{M}_1]^2} - 2\tau \frac{\Cov(\tilde{M}_1\bar{Y}_{11}, \tilde{M_1})\mbbE[\tilde{M}_1\bar{Y}_{11}]}{\mbbE[\tilde{M}_1]^3} + \tau \frac{\Var(\tilde{M}_1)\mbbE[\tilde{M}_1\bar{Y}_{11}]^2}{\mbbE[\tilde{M}_1]^4}. 
\end{align*}
Choosing $\tau=1$ yields the desired result.
This completes the proof for the consistency of the double-or-nothing
bootstrap estimator.
\end{itemize}

\section{Protocol design}\label{sec:protocol_design}

Many vision and audio datasets comprise hundreds of thousands of instances,
making it computationally infeasible to estimate $\fnmr$ and $\fmr$ 
on all the data. 
In such cases, 
the researcher has to decide on which instance pairs their 
computational resources (i.e., budget)
should be spent on. 
Since different combinations of pairwise comparisons between instances may lead to different estimates of model accuracy, 
dataset designers attach protocols specifying which comparisons
to consider in computations. 
Consequently, a natural question is then:
{\it For a given budget, which instance pairs offer 
the lowest variance estimate of model accuracy?}

\begin{figure*}[t]
    \centering
    \includegraphics[width=0.3\textwidth]{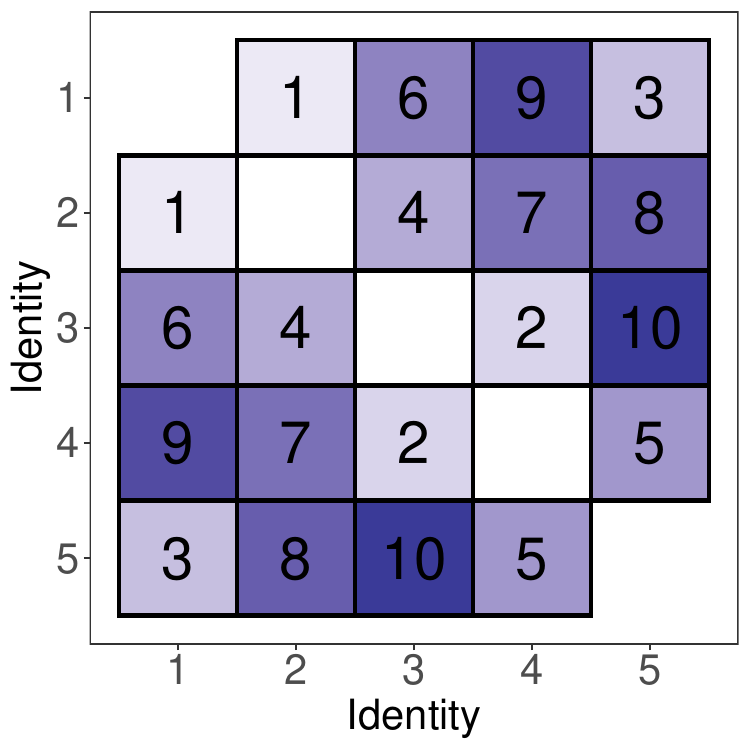}
        \includegraphics[width=0.3\textwidth]{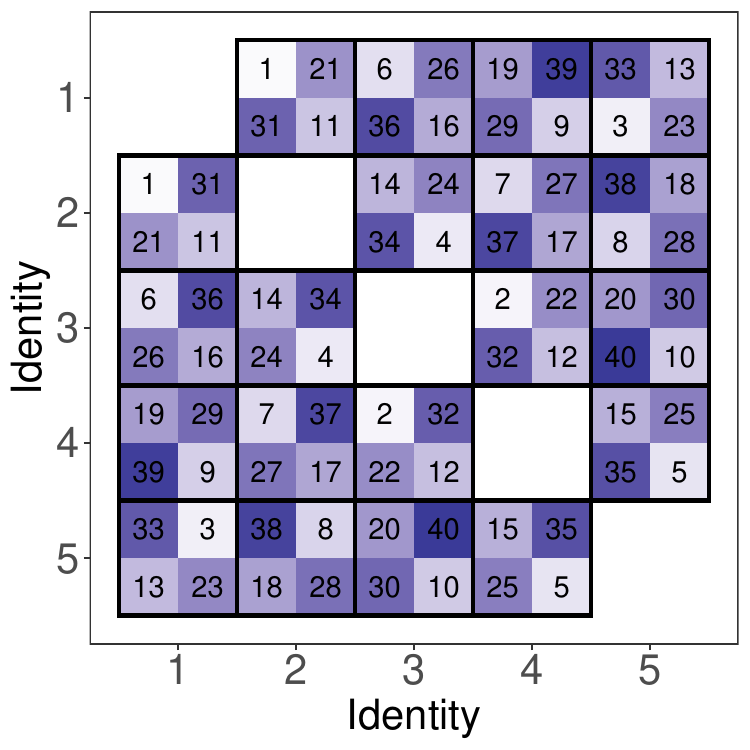}
        \includegraphics[width=0.3\textwidth]{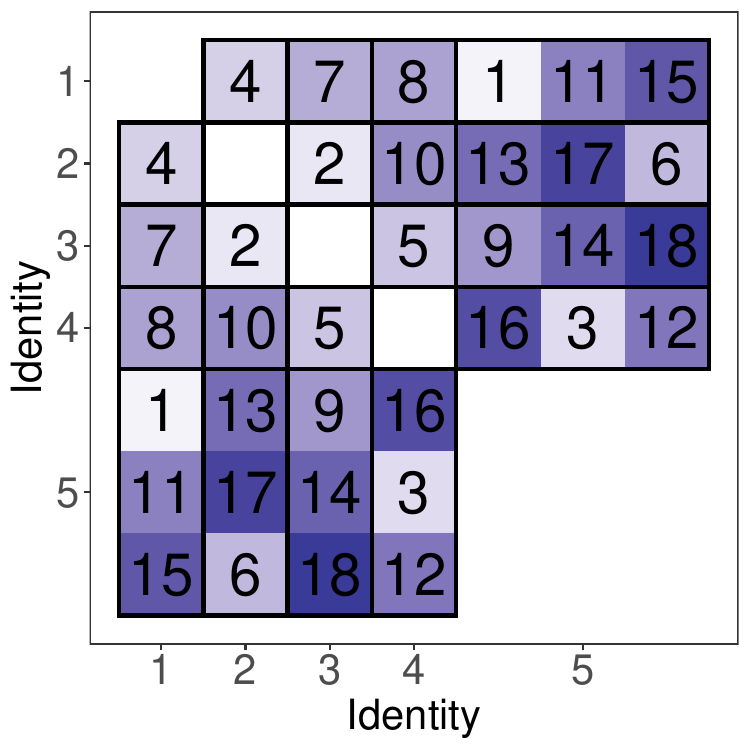}
    \caption{ {\bf Protocol design for selecting identities and instances in $\fmr$ computations.} The selection is based on Algorithm \ref{alg:protocol_design}. The number in each cell represents the iteration at which the given pair of identities and instances is chosen. Left panel: balanced setting with one instance for each of the five identities. Middle panel: balanced setting with two instances for each of the five identities. Right panel: unbalanced setting with the first four and the fifth identities have one and three instances respectively.  Note that in all scenarios we iterate through all identities before selecting the same identity again. However, while in the balanced setting we pick the combination of identities 1--2 first, in the unbalanced setting we choose identities 1--5 first as the latter has more observations.}
    \label{fig:protocol_design}
\end{figure*}

Based on our theoretical analysis in \Cref{sec:methods} 
(and demonstrated in the empirical results in \Cref{sec:experiments}),
it is clear that the dependence structure
induced by the comparisons
can significantly impact the coverage of
the confidence intervals.
This realization naturally leads to a strategy for protocol design
to tries to maintain the independence structure between comparisons.
For simplicity, 
consider the computation of $\fmr$ on a sample where each identity has only one instance. 
Assume that a budget of $B\leq G(G-1)$ comparisons is available,
and let $B=\sum_{i=1}^G \sum_{j\neq i} b_{ij}$ with  $b_{ij}=b_{ji}=1$ when $\bar{Y}_{ij}$ enters $\fmr$ computations and $0$ otherwise. 
Minimizing the variance of the estimated $\fmr$ under budget constraints boils down to solving the following problem: 
\begin{align}
\label{eq:objective_variance_protocol}
    \argmin_{b_{12}, \dots, b_{G(G-1)}} \sum_{\substack{i, j, k=1\\j \neq i\\k \neq i, j}}^G b_{ij} b_{ik} \text{ s.t. } \sum_{\substack{i, j=1\\i \neq j}}^G b_{ij} = B.
\end{align}
When $B \leq \lfloor G/2\rfloor$, one can choose instance pairs that are independent, e.g., $\bar{Y}_{12}, \bar{Y}_{34}$, etc. When $B >  \lfloor G/2\rfloor$, the objective in \eqref{eq:objective_variance_protocol} is minimized when the comparisons share as few instances as possible with each other. An approach to choose the terms to include in the $\fmr$ computations is as follows: At each of the $B$ iterations, select the observation that minimizes the objective evaluated using the allocation resulting from the previous iteration.

\begin{algorithm}
\caption{Protocol Design Strategy to Select Identities Combinations (IDs) for $\fmr$ Estimation in the Balanced Setting}
\label{alg:protocol_design}
\begin{algorithmic}[1]
\Require \texttt{budget} $> 0$, \texttt{data} $= \{\texttt{id}: \texttt{instances}\}$
\State Initialize $\texttt{IDCombinations}$ to empty and $\texttt{IDVisits}$ to priority queue for number of ID visits with IDs present in $\texttt{data}$
\While{$\texttt{budget}>0$}
\State Retrieve candidate IDs with lowest number of visits from $\texttt{IDVisits}$
\State Sort candidate IDs in decreasing order according to the number of instances
\State Iterate through candidate IDs and find first unused pair
\State Update $\texttt{IDCombinations}$, $\texttt{IDVisits}$, and $\texttt{budget}$
\EndWhile
\Ensure Set of ID combinations
\end{algorithmic}
\end{algorithm}

Algorithm \ref{alg:protocol_design} outlines the proposed approach 
for selecting the combinations of identities to be included in the $\fmr$ estimation for the balanced setting. 
We start by creating all possible combinations of identities 
from which we will draw the instances to be considered. 
At each iteration, we use a priority queue to retrieve the identity candidates with the lowest number of visits. 
These candidates are sorted to ensure that those with a larger number of instances are visited first, which helps minimize the number of times a given instance will be reused in the estimation. Note that in the balanced setting, the last step is not necessary. If the total budget exceeds $\lfloor G(G-1)/2\rfloor $, we can iterate through the combinations yielded by Algorithm \ref{alg:protocol_design}. Once the combinations of identities are available, we follow a similar strategy for selecting the pairs of instances within each pair of identities. Figure \ref{fig:protocol_design} describes three examples of protocols resulting from applying this strategy.
For $\fnmr$ estimation, 
we follow a similar idea. 
We first iterate through the identities starting with those with the largest number of instances. 
We then use 
Algorithm \ref{alg:protocol_design}
to select the comparisons within each identity. 

Finally, a brief note about computations of error metrics and the associated uncertainties on massive datasets under computational constraints. The proposed strategy for protocol design can be applied to handle estimation in these settings. 
This involves selecting a fixed number of instance pairs using the protocol design, estimating the error metrics on these pairs, and then using Wilson or bootstrap methods to obtain confidence intervals. 
By following this approach, 
one can obtain reliable estimates of error metrics and their uncertainties
while minizing computational costs.

\section{Additional numerical experiments}\label{sec:experiments_supplement}

In this section, we present additional experimental details and results,
supplementing those in presented in
\Cref{sec:intro} and \Cref{sec:experiments}.

\subsection{Analysis of interval widths}

\begin{figure}[!ht]
    \centering
        \includegraphics[width=0.95\textwidth]{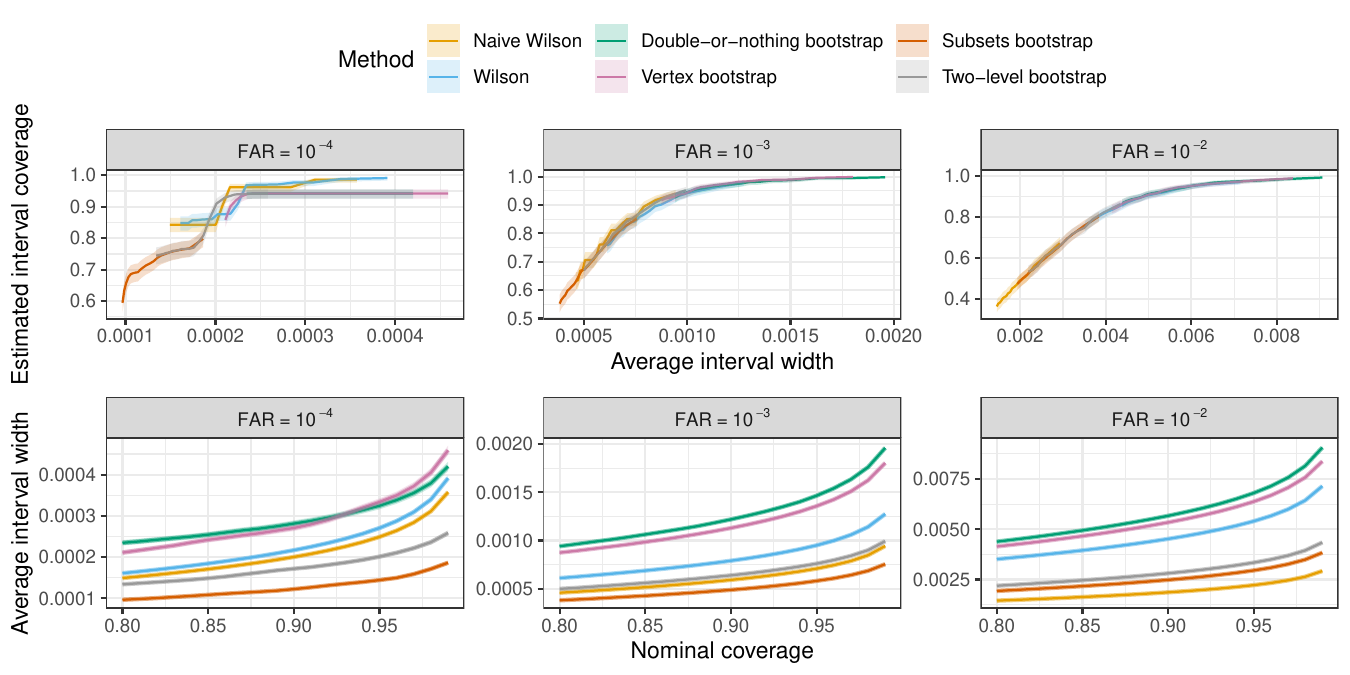}
    \caption{{\bf Estimated coverage versus average width (top) and average width vs. nominal coverage (bottom) for $\fmr$ intervals on synthetic data.} Data contain $G=50$ with $M=5$ instances each. Colored lines and shaded regions indicate estimated coverage and corresponding 95\% naive Wilson or Wald confidence intervals for the different methods.}
    \label{fig:fmr_width}
\end{figure}

The discussion in the main paper has focused on interval coverage and has only briefly mentioned width. In our experiments, we found that methods that yield intervals with higher coverage also generally presented larger widths, as we would expect in case of statistics that are asymptotically normal (see \Cref{prop:normality}). Figure \ref{fig:fmr_width} shows the relationship between estimated coverage, average width, and nominal coverage for $\fmr$ intervals with $G=50$ and $M=5$ using the setup described in \Cref{sec:synthetic} (see Figure \ref{fig:balanced_coverage} for estimated vs. nominal coverage). In case of $\fmr=10^{-3}, 10^{-4}$, a given estimated coverage corresponds to the same interval width across all methods. This indicates that recalibrating the nominal coverage (e.g., increasing the nominal level $1-\alpha$ for the subsets or two-level bootstrap to achieve intervals with coverage $1-\alpha_{\text{target}}$) for any of the methods will not yield intervals with the target coverage but with smaller width. 

\editedinlinetwo{
\subsection{Pointwise intervals for the ROC}

We evaluate the coverage of pointwise confidence intervals for the ROC on MORPH data. 
The experimental setup follows the description of \Cref{sec:experiments}. 
The vertex bootstrap performs similarly to the double-or-nothing procedure and thus, for the sake of simplifying the presentation of the results, it is omitted from the discussion. 
\Cref{fig:synthetic_roc} shows estimated coverage as a function of nominal coverage for the reviewed methods at different levels of $\fmr$. 
Consistently with the discussion of Section \ref{sec:pointwise_roc}, we observe that Wilson intervals achieve coverage that is higher than nominal across all $\fmr$ levels. 
While the overcoverage may be expected for low values of $\fnmr$ (e.g., see the results in \Cref{fig:balanced_coverage}), the overestimation is present albeit it is lower for larger values of $\fnmr$. 
For low $\fnmr$, we also observe that the version of the double-or-nothing bootstraps which employ ROC curves smoothed using log-normal distributions to model the scores perform better than their counterparts. 
This is suggestive of the benefits of imposing smoothness assumptions. When $\fnmr$ is large enough, the bootstraps perform similarly. 

\begin{figure}[!ht]
    \centering
    \includegraphics[width=0.95\textwidth]{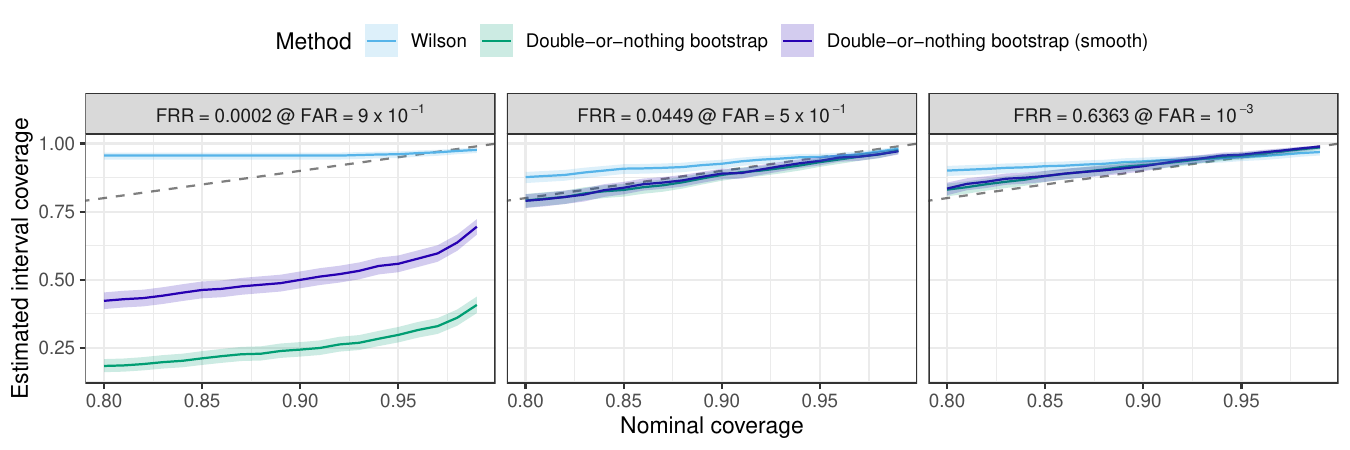}
    \caption{
    {\bf Estimated coverage versus nominal coverage of confidence intervals for $\fnmr$@$\fmr$ on MORPH data.} 
    Samples were generated by resampling $G=50$ identities from the original dataset without replacement.
    }
    \label{fig:synthetic_roc}
\end{figure}
}

\editedinlinetwo{
\subsection{Experiments on diverse data types}

Our theoretical analysis applies to any 1:1 matching task.  Here we explore its properties empirically  on real data, with data types and tasks beyond 1:1 face verification.  
In particular, we investigate the performance of our methods in the following tasks: 

\medskip

\begin{itemize}
\item \textbf{1:1 object verification.}
Matching images from a randomly sampled subset of the iNat2021 dataset \citep{van2021benchmarking}. 
The iNat2021 dataset is an image collection specifically curated for species recognition, featuring over 10,000 different species. The matching tasks is to recognize whether the animals in two different images belongs to the same species. For this purpose, we extract feature representations obtained via CLIP \citep{radford2021learning}.

\smallskip

\item \textbf{1:1 speaker verification.}
We use a large dataset of voice recordings corresponding to multiple speakers. For the speaker verification task, we extract the embeddings of the audio recordings using an \texttt{ECAPA-TDNN} pretrained model \cite{speechbrain, desplanques2020ecapa}. 

\smallskip

\item \textbf{1:1 topic verification.}
We aim to detect whether two text paragraphs are related to the same topic. For this purpose, we use a subset of the Amazon review dataset \citep{ni2019justifying}, comprising product information and corresponding Amazon reviews. Specifically, we focus on identifying if two reviews pertain to the same product. Classification is performed using text embeddings generated by \texttt{BAAI/bge-smal-en-v1.5} \citep{bge_embedding}. 
\end{itemize}


\begin{figure}[t]
    \centering
    \includegraphics[width=0.95\textwidth]{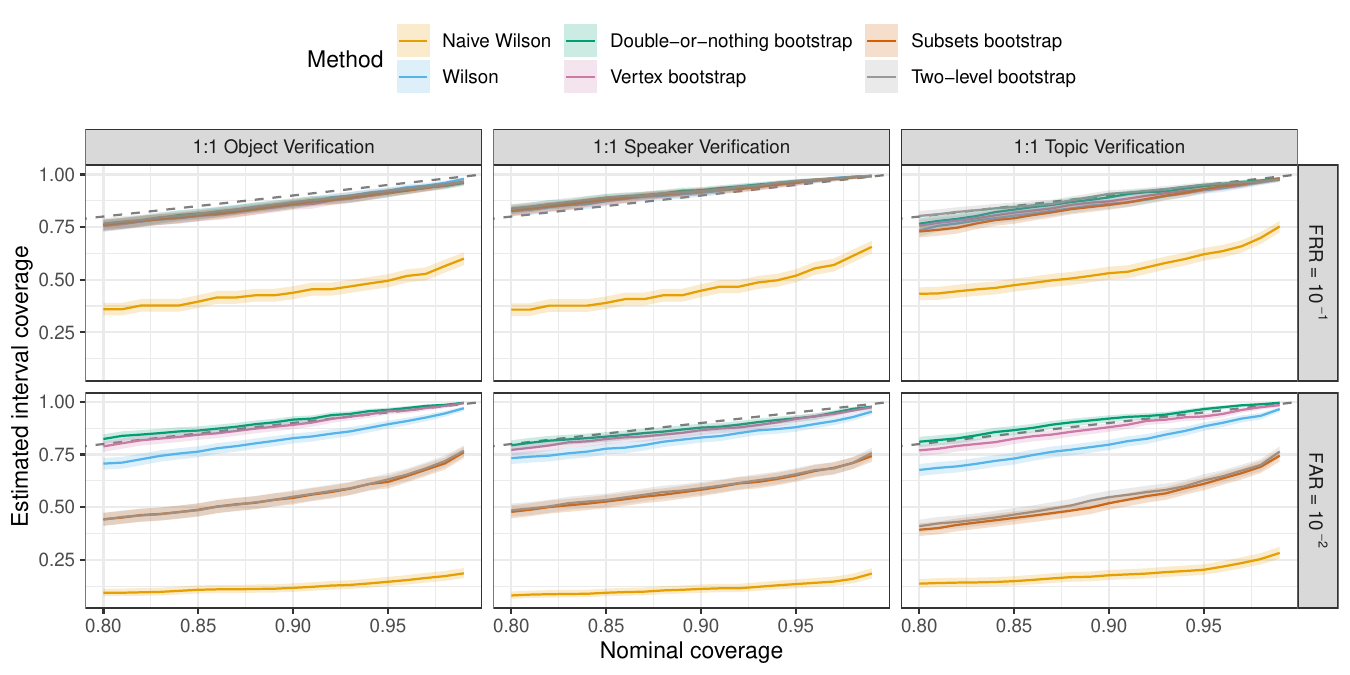}
    \caption{ {\bf Estimated interval coverage vs. nominal coverage for $\fnmr=10^{-1}$ (top) and $\fmr=10^{-2}$ (bottom) on 1:1 object, speaker, and topic verification tasks respectively.} Samples were generated by resampling $G=50$ identities without replacement from the original dataset. 
    }\label{fig:multiple_unbalanced}
\end{figure}

For all datasets, our experimental framework follows the same setup of Section \ref{sec:experiment_morph}, using $G=50$ identities. 
\Cref{fig:multiple_unbalanced} shows show how the coverage of the confidence intervals for $\fnmr=10^{-1}$ and $\fmr=10^{-2}$, constructed using the reviewed methods, varies with nominal coverage. The results are consistent with our empirical findings of \Cref{sec:experiments}: For $\fnmr$, all methods other than naive Wilson tend to cover approximately at the right level. For $\fmr$, the Wilson intervals, as well as the vertex and double-or-nothing bootstrap intervals, achieve coverage close to nominal. The other methods severely undercover. 
}



\end{appendices}

\end{document}